\documentclass[11pt,a4paper]{article}

\usepackage[utf8]{inputenc}
\usepackage{CJKutf8}

\usepackage{amsmath,amssymb,amsthm}

\usepackage{graphicx}
\usepackage{booktabs}
\usepackage{multirow}
\usepackage{subcaption}
\usepackage{rotating}
\usepackage{graphicx}
\usepackage{subcaption}
\graphicspath{{Figure/}}

\usepackage{xcolor}

\usepackage{hyperref}
\hypersetup{
	colorlinks=true,
	linkcolor=blue,
	citecolor=blue,
	urlcolor=blue
}

\usepackage[margin=2.5cm]{geometry}

\usepackage{enumitem}
\usepackage{float}

\DeclareUnicodeCharacter{FF08}{(}
\DeclareUnicodeCharacter{FF09}{)}
\DeclareUnicodeCharacter{2013}{--}
\DeclareUnicodeCharacter{2014}{---}

\title{\textbf{Frequency-Guided Deformable Networks for Continuous Phase Alignment}}

\author{
	Wangye Jiang\textsuperscript{3,$\dagger$} (wjiang18@CougarNet.UH.EDU),\\
	Haoming Yang\textsuperscript{2,$\dagger$} (2222441012@stu.jit.edu.cn),\\
	Jian Xu\textsuperscript{1} (sopher@szut.edu.cn),\\
	Jingya Zhang\textsuperscript{1,$*$} (zhangjy0611@163.com)
}

\date{
	\textsuperscript{1}School of Electronic and Information Engineering, Suzhou University of Technology, Suzhou 215500, Jiangsu, China\\
	\textsuperscript{2}School of Software Engineering, Jinling Institute of Technology, Nanjing 211169, Jiangsu, China\\
	\textsuperscript{3}Cullen College of Engineering, University of Houston, Houston, TX 77204, USA
}

\begin{document}
	\begin{CJK*}{UTF8}{gbsn}
		
		\maketitle

		\begin{abstract}
			The core of time series analysis lies in effectively modeling the physical laws within complex signals. Existing Transformer and Convolution Neural Network (CNN) architectures are often constrained by insufficient temporal inductive bias, restricted frequency extraction capabilities, or weak local phase alignment. To this end, this paper proposes Adaptive Network Based on Cascaded Harmonic Offset Routing (ANCHOR), an Adaptive Network based on Cascaded Harmonic Offset Routing. The model utilizes the Real Fast Fourier Transform (RFFT) to extract explicit dominant periods, injecting them as physical anchors into the dilation operators of multi-branch deformable convolutions. This guides the adaptive optimization of sampling locations in the time domain, achieving synergistic modeling of macroscopic periodic priors and microscopic geometric deformations. Furthermore, to address the quantization errors and picket-fence effects introduced by the discrete RFFT, this paper imports a continuously differentiable 1D Gaussian Radial Basis Function interpolation operator to replace traditional linear interpolation. This maintains the differentiability of the interpolation process and enhances the accuracy of sub-pixel phase compensation. Additionally, ANCHOR introduces an asymmetric routing mechanism and orthogonal channel partitioning to dynamically balance the extraction weights between high-energy strong signals and low-energy weak features. Multi-task benchmark experiments demonstrate that ANCHOR achieves the best or solid performance in short-term forecasting, anomaly detection, and time series classification tasks. Code is available at \url{https://github.com/Jwy-EE/Anchor_pub}
		\end{abstract}
		
		\section{Introduction}
		\label{sec:intro}
		
		Time series analysis encompasses time series forecasting, classification, and recognition. Although their task objectives vary, their essence universally relies on the deep deconstruction of the underlying physical laws within complex time series signals. Currently, this field serves as a critical infrastructure for system decision-making and robustness evaluation \cite{liang2024foundation,qiu2026rethinking}. In climate modeling and power dispatching, high-precision time series forecasting is a prerequisite for resource allocation optimization and disaster early warning \cite{wang2024timexer,wu2022timesnet}; in the fields of the Industrial Internet of Things (IIoT) and cybersecurity, real-time anomaly detection serves as a critical means to identify non-stationary deviations, ensuring the operational safety of complex systems \cite{mahesh2025themis,ahmed2016survey}; while in biomedical signal processing, precise time series classification is the core of achieving automated clinical diagnosis and pathology recognition \cite{langer2025opentslm,goldberger2000components}. Real-world time series data are often intertwined with multi-scale and multi-period patterns. Such periodic characteristics not only dictate forecasting trends but also constitute structural fingerprints for classification and physical baselines for anomaly determination \cite{wu2022timesnet}. Therefore, constructing a backbone network capable of explicitly capturing physical periodic priors and possessing cross-task universality has emerged as a frontier challenge in the current field of time series analysis \cite{wu2022timesnet,gao2024units}.
		
		In early studies, Autoregressive Integrated Moving Average (ARIMA) models established linear modeling norms based on stationarity assumptions and autocorrelation structures through rigorous mathematical derivations \cite{box2015time}. To further capture the dynamic evolution in data, the Holt-Winters exponential smoothing method introduced explicit modeling of trend and seasonal components, demonstrated interpretability in specific scenarios in electrical load and business forecasting \cite{holt2004forecasting,su2022dlinear}. Since the introduction of the Transformer, it has demonstrated strong performance across a wide range of tasks, such as natural language processing \cite{achiam2023gpt,grattafiori2024llama}, speech recognition \cite{radford2023robust,barrault2023seamlessm4t}, and computer vision \cite{peebles2023scalable,kirillov2023segment}. Similarly, the Transformer architecture has focused on challenging domains within time series tasks, such as long-term forecasting tasks, effectively expanded the temporal receptive field utilizing parallel computing mechanisms, exemplified by Informer \cite{zhou2021informer}, Autoformer \cite{wu2021autoformer}, and PatchTST \cite{nie2022time}. Furthermore, in zero-shot forecasting tasks, its capability of large-scale pre-training on massive, heterogeneous multi-domain datasets has resolved many real-world cold-start problems, as seen in TimeGPT-1 \cite{garza2023timegpt}, Chronos \cite{ansari2024chronos}, and Moirai \cite{woo2024unified}. Concurrently, CNN architectures have witnessed a renaissance in time series tasks due to their inherent local feature extraction capabilities and lower parameter costs. TimesNet demonstrated the advantages of convolutions in processing periodic features \cite{wu2022timesnet}; additionally, SCINet and MICN proved the effectiveness of multi-scale convolutions in modeling long-range dependencies \cite{liu2021time,wang2023micn}, and ModernTCN, through a modern pure convolution architecture, has comprehensively surpassed mainstream Transformers in both accuracy and efficiency \cite{luo2024moderntcn}.
		
		However, existing architectures exhibit specific theoretical and practical trade-offs. On the one hand, although Transformers and their derived foundation models have demonstrated powerful global representation capabilities in modeling long-range dependencies, their implicit attention mechanisms struggle to explicitly capture the macroscopic periodic invariants and continuous local phase shifts inherent in time series. Real-world time series signals heavily depend on strict chronological order and continuous local dependencies, whereas the core self-attention mechanism of the Transformer possesses inherent permutation invariance. As pointed out by Zeng et al. \cite{zeng2023transformers}, even the introduction of positional encoding may not fully compensate for this structural loss, occasionally causing performance to trail behind robust linear baselines on certain benchmarks. Recent studies on in-context learning indicate that linear self-attention, while powerful, may face theoretical performance upper bounds comparable to classical linear regression in specific tasks, and can exhibit stability challenges during long-sequence inference \cite{zhou2025transformers}. Additionally, their high parameter capacity typically requires massive training data, posing severe overfitting risks and computational bottlenecks in small-sample scenarios \cite{wen2022transformers}. These observations suggest that directly migrating attention-heavy architectures to time series analysis may incur $O(N^2)$ computational and memory overhead while lacking the precise physical inductive biases necessary for time series data generation at the architectural level. On the other hand, although standard Convolutional Neural Network architectures have achieved widespread success in time-series feature extraction, they still face certain structural limitations when processing non-stationary signals. First, at the spatial sampling level, standard CNNs rely on discrete, equidistant sampling grids. When confronted with minor sub-pixel phase drifts within signals, such rigid grid structures struggle to achieve continuous and precise feature alignment. Second, to accommodate the highly variable frequencies and periodic spans inherent in time series, existing mainstream methods typically employ parallel multi-scale branches. While this redundancy-based strategy effectively expands the feature capture range, it fails to decouple the receptive field from computational weights at the underlying mechanistic level, concurrently incurring high computational overhead \cite{wu2022timesnet,liu2021time,wang2023micn,luo2024moderntcn}.
		
		To compensate for the lack of structural physical priors in the aforementioned backbones, recent works have attempted to explicitly introduce frequency-domain information to assist spatial modeling. For instance, Autoformer \cite{wu2021autoformer} introduces an auto-correlation mechanism based on series periodicity; TimesNet \cite{wu2022timesnet} extracts multiple periods via the Fast Fourier Transform and reshapes 1D sequences into 2D tensors. However, these methods are primarily built upon strict static period assumptions. When aggregating frequency components or performing 2D spatial folding, such fixed mechanisms lack the structural flexibility to accommodate continuous local phase shifts. When local phase shifts occur in real-world data, fixed spatial reshaping leads to feature misalignment and over-smoothing, thereby limiting the model's capability to capture microscopic dynamic variations.
		
		To address the aforementioned issues, this paper proposes the ANCHOR model. It is argued that macroscopic period priors should not be utilized merely for rigid spatial folding, but are more appropriate as sampling reference baselines along the temporal axis. Specifically, the RFFT is utilized to extract the dominant periods of the signal, injecting them as prior anchors directly into the underlying dilation operators of multi-branch deformable convolutions. This integration mechanism establishes a macroscopic sampling baseline while allowing the convolutional operators to adaptively adjust sampling positions within the continuous 1D time domain, achieving continuous phase alignment.
		
		Nevertheless, implementing this mechanism encounters a computational contradiction between discrete priors and continuous phases: the periods extracted by the discrete RFFT are integer sampling points, whereas the true phase offsets are typically continuous real numbers. When the model attempts to compensate for such sub-pixel mismatches, the bilinear interpolation employed in standard deformable convolutions (e.g., DCNv4 \cite{xiong2024efficient}) exhibits limitations. Due to its piecewise linear nature, local gradient truncation frequently occurs when optimizing continuous phase offsets, which adversely affects model convergence. To resolve this, an infinitely differentiable 1D Gaussian radial basis interpolation operator is introduced. Furthermore, to address the signal-to-noise ratio imbalance in multi-scale feature extraction, this paper proposes an asymmetric routing strategy of "small kernels to preserve high-energy strong signals, and large kernels to integrate low-energy weak features," effectively balancing the extraction weights of features across different frequency bands. Coupled with a progressive residual cascade architecture, ANCHOR effectively alleviates time-domain entanglement, achieving a balance between the sharpness of strong signals and the noise floor of weak signals. Empirical evaluations demonstrate that ANCHOR exhibits robust performance across multi-task benchmarks. The main contributions of this paper are as follows:
		
		\begin{enumerate}[leftmargin=*]
			\item \textbf{Revealing prior-guided deformation sparsity and a cross-domain phase compensation mechanism based on Gaussian kernels.} This work introduces frequency-domain harmonics as physical anchors for spatial geometric deformations in time series deep learning. By injecting the dominant periods extracted by RFFT into the underlying dilation operators of deformable convolutions, the model establishes a macroscopic sampling baseline; furthermore, by introducing a continuously differentiable 1D Gaussian radial basis interpolation operator, effectively overcomes the gradient truncation defect of traditional linear interpolation under sub-pixel displacements. This mitigates the inherent quantization errors caused by the discrete picket-fence effect of RFFT in the time domain.
			\item \textbf{Creating a backbone network based on an energy-compensating frequency-space adaptation criterion.} Addressing the signal-to-noise ratio imbalance of real-world non-stationary time series signals, this paper constructs a hierarchical progressive cascade architecture based on channel partitioning, founded on an asymmetric energy-compensating routing mechanism of ``small kernels to preserve strong signals, and large kernels to integrate weak signals.'' This architecture overcomes the issue where strong signals are over-smoothed by large kernels and weak signals are interfered with/corrupted by noise in complex environments, achieving deep decoupling of macroscopic laws and microscopic perturbations at the source of feature circulation.
			\item \textbf{Through systematic multi-task benchmark evaluation, this study validates the cross-domain effectiveness of the ANCHOR framework.} The architecture yields robust quantitative results across diverse time series modeling paradigms: it establishes a generalizable baseline in short-term forecasting, achieving an OWA of 0.766 and SMAPE of 13.032 on the M4-Yearly benchmark, with a highly competitive average OWA of 0.836 across all subsets; it ensures precise fault localization in anomaly detection, attaining an AUPRC of 0.8112 on the UCR archive; and it demonstrates excellent discriminative performance on highly challenging sequence classification datasets.
		\end{enumerate}
		
		\section{Methodology}
		
		\subsection{Explicit Frequency-Domain Physical Prior Extraction}
		
		Given a historical multivariate time series tensor $\mathcal{X}\in\mathbb{R}^{B\times C\times L}$, where $B$ is the batch size, $C$ is the number of channels, and $L$ is the lookback window length, we perform a RFFT on the input tensor along the temporal dimension:
		\[
		\mathcal{X}_{\text{freq}}=RFFT(\mathcal{X},dim=-1)\in\mathbb{C}^{B\times C\times(\lfloor L/2\rfloor+1)}.
		\]
		
		To filter out channel-specific local noise, we average the spectral magnitudes across the batch and channel dimensions to obtain a globally unified spectral energy distribution vector $a\in\mathbb{R}^{\lfloor L/2\rfloor+1}$. For a frequency point with index $f$, its energy magnitude is defined as:
		\[
		a(f)=\frac{1}{B\cdot C}\sum_{b=1}^{B}\sum_{c=1}^{C}|\mathcal{X}_{b,c,f}|.
		\]
		
		After removing the DC component by setting $a(0)=0$, we select the top-$K$ frequency indices with the highest energy to form the set $\mathcal{F}_{\text{top}}=\{f_{1},f_{2},...,f_{K}\}$. The mapped physical period set $\mathcal{P}=\{T_{1},T_{2},...,T_{K}\}$ is defined as:
		\[
		T_{k}=\left\lfloor\frac{L}{f_{k}}\right\rfloor,\quad \forall k\in\{1,2,...,K\}.
		\]
		
		\subsection{Frequency-Guided Deformable Module (FGDM)}
		\subsubsection{Continuous Phase Alignment via Prior-Guided Deformation}
		As previously mentioned, the introduction of RFFT effectively provides the model with global macroscopic harmonic priors. However, the evolution of physical signals in the real world occurs in a continuous time domain. Directly applying the discrete Fourier transform inevitably triggers the picket-fence effect. This means that the explicit physical period $T_{k}$ extracted by the front-end module is necessarily a quantized discrete integer, which contains inherent rounding errors. Such an error-prone discrete period cannot be aligned with absolute precision to the true continuous physical phase. On the one hand, zero-padding fundamentally only provides a global, static, and precise period, rendering it completely incapable of handling the local and dynamic phase drifts that are ubiquitous in realistic non-stationary signals. On the other hand, although introducing a flat-top window can compensate for amplitude errors, it severely broadens the spectral main lobe, directly reducing the geometric sharpness of the frequency resolution.
		
\begin{figure}[htbp]
			\centering
			\includegraphics[width=0.88\linewidth]{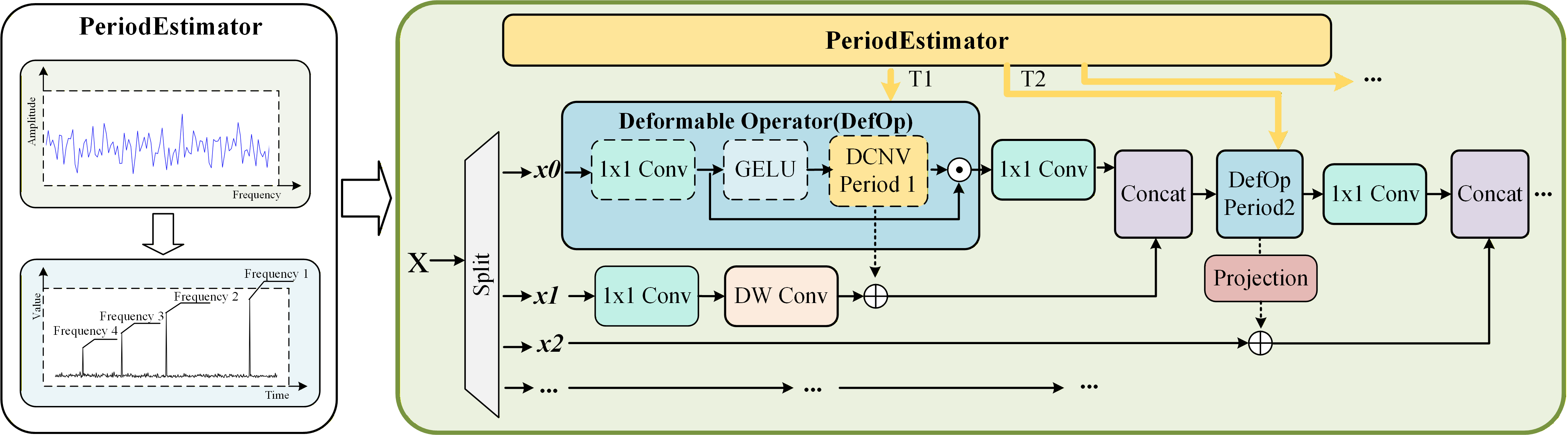}
			\caption{Structure diagram of FGDM}
			\label{fig:fgdm}
		\end{figure}

		To avoid the limitations of static compensation by zero-padding and frequency resolution destruction by flat-top windows found in traditional digital signal processing, this paper proposes a novel cross-domain error compensation architecture. This architecture does not require lossy repair within the frequency domain, but rather uses only the discrete period $T_{k}$ with quantization errors as a coarse-grained spatial anchor. As shown in Fig.~\ref{fig:fgdm}, this cross-domain compensation mechanism takes the discrete period $T_{k}$ as a coarse spatial anchor and recovers the continuous time-domain phase underlying the signal through continuous deformable sampling. To this end, a one-dimensional deformable convolution is introduced with continuous spatial interpolation capabilities across domains is first performed. Unlike the original DCNv4, a one-dimensional temporal adaptation on DCNv4. Secondly, to adapt to 1D variable receptive field modeling, the front end of the convolution adopts an explicit convolution execution method, dynamically setting padding and dilation during the forward process, while replacing bilinear interpolation with Gaussian radial basis interpolation, allowing it to spontaneously complete the alignment and reconciliation of continuous phases in the time domain.
		
		For a given local one-dimensional feature sequence $x\in\mathbb{R}^{L}$, under standard convolution, the relative sampling grid point set for a convolution kernel of size $S_{k}$ is $\mathcal{G}=\{-\lfloor S_{k}/2\rfloor,...,\lfloor S_{k}/2\rfloor\}$. After injecting the physical period $T_{k}$, the actual continuous spatial absolute coordinate $p_{n}$ of the $n$-th sampling point ($n\in\mathcal{G}$) is reconstructed as:
		\[
		p_{n}=p_{0}+T_{k}\cdot n+\Delta p_{n},
		\]
		where $\Delta p_{n}$ is the continuous sub-pixel offset spontaneously learned by the network, and $p_{0}$ is the central sampling coordinate.

		\subsubsection{ $C^\infty$ Smooth Sub-pixel Interpolation via Gaussian RBF}
		Considering that the reconstructed sampling coordinate $p_{n}$ exists in a continuous real number space, the traditional bilinear interpolation only possesses $C^{0}$ continuity due to its piecewise linear mathematical nature. During the backpropagation process, when the offset predicted by the network deviates from the true physical phase by more than one discrete sampling grid, its local derivative will experience gradient truncation, which in turn causes the optimization trajectory to easily fall into a local minimum. To address this defect, a one-dimensional Gaussian radial basis interpolation with infinite differentiability is introduced:
		\[
		x(p_{n})=\sum_{q\in\Omega}x(q)\cdot
		\frac{\exp\left(-\frac{(p_{n}-q)^{2}}{2\sigma^{2}}\right)}
		{\sum_{q'\in\Omega}\exp\left(-\frac{(p_{n}-q')^{2}}{2\sigma^{2}}\right)},
		\]
		where $\Omega$ represents the local receptive window centered on the continuous target point $p_{n}$, $q$ is the discrete integer grid coordinate traversed within the window, and $\sigma$ is the scale hyperparameter controlling the distribution width of the kernel function. This operator ensures continuous derivative flow across the entire loss landscape by constructing a smooth exponentially decaying gravitational field around the sampling point. Even in cases with large initial deviations, the broad gradient receptive field of the Gaussian kernel can still provide stable traction, guiding the sampling point to accurately converge toward the true physical phase.
		
		While the topological advantages of the Gaussian RBF over traditional methods are conceptually intuitive, a rigorous mathematical analysis of their backpropagation dynamics is essential to expose the exact mechanics of gradient propagation. To substantiate the "gradient truncation" crisis inherent in $C^0$ bilinear interpolation and strictly contrast it with the $C^\infty$ continuous gradient flow of our proposed Gaussian RBF, it is essential to delve into the derivation of their respective local feature gradients $\frac{\partial x(p_n)}{\partial p_n}$
		
		The standard 1D bilinear interpolation is first revisited to expose its flaw under fractional offsets. In the local feature grid, let the continuous sampling point coordinate be $p_n$. When $p_n$ falls between two discrete integer indices $q_L = \lfloor p_n \rfloor$ and $q_R = q_L + 1$, the 1D bilinear interpolation is defined as:
		
		$$
		x(p_n) = x(q_L) \cdot (1 - d) + x(q_R) \cdot d
		$$
		
		where the fractional offset is defined as $d = p_n - q_L$, satisfying the boundary condition $0 \le d < 1$.To compute the partial derivative of the loss function $\mathcal{L}$ with respect to the learnable offset $\Delta p_n$, we apply the chain rule:
		
		$$
		\frac{\partial \mathcal{L}}{\partial \Delta p_n} = \frac{\partial \mathcal{L}}{\partial x(p_n)} \cdot \frac{\partial x(p_n)}{\partial p_n} \cdot \frac{\partial p_n}{\partial \Delta p_n}
		$$
		
		Given that the local linear derivative of the minute offset is $\frac{\partial p_n}{\partial \Delta p_n} = 1$, the problem strictly converges to solving the local feature gradient $\frac{\partial x(p_n)}{\partial p_n}$. When the point is continuously and minutely perturbed within the open interval $(q_L, q_R)$, the base anchor $q_L$ remains constant. Taking the partial derivative with respect to the continuous variable $p_n$ yields:
		
		$$
		\frac{\partial x(p_n)}{\partial p_n} = x(q_R) - x(q_L)
		$$
		
		When employing a Gaussian kernel to expand the local receptive field $\Omega$ and achieve globally smoothed resampling, the function mapping of the forward pass becomes:
		
		$$
		x(p_n) = \frac{\sum_{q\in\Omega} x(q) \cdot \exp\left(-\frac{(p_n-q)^2}{2\sigma^2}\right)}{\sum_{q'\in\Omega} \exp\left(-\frac{(p_n-q')^2}{2\sigma^2}\right)}
		$$
		
		where $\sigma$ is the scale hyperparameter controlling the distribution width of the kernel function.To achieve variable decoupling and simplify the differential equation structure, the unnormalized weight shared by the numerator and denominator is extracted, defined as $w_q(p_n) = \exp\left(-\frac{(p_n-q)^2}{2\sigma^2}\right)$, and the normalization factor integration term $W(p_n) = \sum_{q'\in\Omega} w_{q'}(p_n)$. First, applying the chain rule to $w_q(p_n)$:
		
		$$
		\frac{\partial w_q(p_n)}{\partial p_n} = \frac{q - p_n}{\sigma^2} w_q(p_n)
		$$
		
		Let the numerator's feature-weighted term be $S(p_n) = \sum_{q\in\Omega} x(q) w_q(p_n)$. Using the quotient rule to differentiate the overall dimensionless interpolation function, and substituting the aforementioned differential fundamentals:
		
		$$
		\frac{\partial x(p_n)}{\partial p_n} = \frac{S'(p_n) W(p_n) - S(p_n) W'(p_n)}{[W(p_n)]^2} = \frac{S'(p_n)}{W(p_n)} - x(p_n) \frac{W'(p_n)}{W(p_n)}$$Substituting the local derivative formulas into the fully expanded expression, isolating the constant positive scalar $\frac{1}{\sigma^2}$, and further defining the actual normalized weight as $\alpha_q(p_n) = \frac{w_q(p_n)}{W(p_n)}$, the analytical solution is obtained for the backpropagation gradient of the Gaussian interpolation through algebraic simplification:
		$$\frac{\partial x(p_n)}{\partial p_n} = \frac{1}{\sigma^2} \sum_{q\in\Omega} \alpha_q(p_n) \cdot (q - p_n) \cdot [x(q) - x(p_n)]
		$$
		
		This rigorous derivation is not only mathematically self-consistent but also physically explains the dynamic mechanism of this interpolation: the gradient update is essentially a mean-shift vector jointly weighted by the relative spatial distance $(q - p_n)$ and the relative feature difference $[x(q) - x(p_n)]$, which effectively circumvents local gradient explosion.In the differentiable spatial sampling architecture, the core advantage of Gaussian interpolation lies in fundamentally reshaping the high-dimensional error surface of backpropagation. Compared to the "gradient dead zones" caused by the local truncation inherent to the $C^0$ continuity of bilinear interpolation, the Gaussian radial basis function constructs a strictly $C^\infty$ infinitely differentiable smooth global gravitational field. This ensures that even if the continuous prediction coordinate severely deviates from the target feature, distant discrete anchors can still provide continuous and directional gradient momentum through long-tail non-zero weights. Consequently, this mechanism mathematically eliminates numerical oscillations during the optimization process, significantly enhancing the global convergence of the spatial offset parameters.
		
		\subsubsection{Orthogonal Subspace Partitioning and Cascaded Routing}
		To introduce this mechanism with lower parameter volume and achieve efficient fusion of multi-scale features, the FGDM plug-and-play module is proposed. Based on the inspiration of the RFA module for image processing in UniconvNet \cite{wang2025uniconvnet}, for the $l$-th network layer, given the input feature $\mathcal{X}^{(l)}\in\mathbb{R}^{B\times C\times L}$, its complete feature evolution can be strictly defined by the following steps. Given the input feature tensor $\mathcal{X}^{(l)}\in\mathbb{R}^{B\times C\times L}$ of the $l$-th layer, in order to construct a multi-path feature route, the model first executes the channel splitting operator $\Gamma(\cdot)$ to project the original feature space into $N$ orthogonal subspace partitions:
		\[
		\{x_{i}\}_{i=0}^{N-1}=\Gamma(\mathcal{X}^{(l)}),\quad x_{i}\in\mathbb{R}^{B\times\frac{C}{N}\times L}.
		\]
		
		Let $\mathcal{D}(\cdot;K,T)$ be a Deformable Operator(DefOp) configured with a specific kernel size $K$ and physical period $T$. A sequence of monotonically increasing receptive field configurations is defined$\mathcal{K}=\{K_{1},K_{2},...,K_{N-1}\}$ and the corresponding physical period prior set $\mathcal{P}=\{T_{1},T_{2},...,T_{K}\}$. To physically motivate the asymmetric routing, the periods in $\mathcal{P}$ are strictly sorted in descending order of their spectral energy. In the routing mechanism, high-energy dominant periods are assigned to smaller receptive fields to maintain high temporal resolution and precise phase localization of primary oscillations. Conversely, low-energy components are routed to larger receptive fields, which function as temporal low-pass filters to smooth out high-frequency noise and extract global background trends. For the cascade step $i\in\{1,...,N-1\}$, the progression of multi-scale frequency-space features can be abstracted into the recursive state equation:
		\[
		y_{i}=
		\left[
		\phi_{ci}(x_{i})+\mathcal{D}_{i}(y_{i-1};K_{i},T_{i})
		\parallel
		\mathcal{D}_{i}(y_{i-1};K_{i},T_{i})\odot\phi_{vi}(y_{i-1})
		\right].
		\]
		
		Here $x_{i}$ is the static original feature partition, and $y_{i}$ is the dynamic cumulative feature representing the $i$-th stage incorporating the preceding multi-scale frequency-space information. $\phi_{ci}(\cdot)$ and $\phi_{vi}(\cdot)$ denote $1\times1$ pointwise convolutions used for dimensional alignment and gated mask generation at stage i, respectively. $\odot$ denotes element-wise multiplication, denotes element-wise multiplication, and $[\cdot\parallel\cdot]$ denotes channel-wise concatenation.A
		
		FGDM can be put into any model as a plug-and-play module. From the above description, it can be seen that its native orthogonal channel partitioning strategy precisely provides mutually isolated physical subspaces for multi-scale frequency extraction. At the same time, thanks to its channel partitioning processing strategy, compared with traditional multi-branch full-channel networks, this instantiated model greatly suppresses computational complexity and parameter redundancy while achieving multi-scale frequency-space fusion. For a traditional multi-branch full-channel network, given an input feature with sequence length $L$ and channel number $C$, assume the architecture contains $N-1$ parallel branches with receptive field configurations $\mathcal{K}=\{K_{1},K_{2},...,K_{N-1}\}$. Since each branch densely maps the complete feature dimensions, the general theoretical complexity of its core spatial operations is:
		\[
		Cost_{\text{Baseline}}=\mathcal{O}\!\left(\sum_{i=1}^{N-1}L\cdot C\cdot C\cdot K_{i}\right)=
		\mathcal{O}\!\left(L\cdot C^{2}\cdot\sum_{i=1}^{N-1}K_{i}\right).
		\]
		
		In contrast, under the orthogonal subspace partitioning strategy, the input tensor is first equally divided into $N$ mutually independent subspaces, and then an $N-1$ level progressive feature cascade is performed. In the $i$-th stage of evolution, the input and output channel dimensions participating in the mapping increase synchronously to $i\cdot\frac{C}{N}$. Therefore, its single-layer computational volume follows a quadratic growth law, and the overall spatial theoretical complexity can be abstracted as:
		\[
		Cost_{\text{Spatial}}=\mathcal{O}\!\left(\sum_{i=1}^{N-1}L\cdot\left(i\cdot\frac{C}{N}\right)^{2}\cdot K_{i}\right)=
		\mathcal{O}\!\left(L\cdot\frac{C^{2}}{N^{2}}\cdot\sum_{i=1}^{N-1}i^{2}K_{i}\right).
		\]
		
		Furthermore, the model only performs a single RFFT global physical period prior extraction on the original variable in the input stage, and its complexity $\mathcal{O}(C_{in}\cdot L\log L)$ accounts for an extremely small proportion and can be ignored. Therefore, the channel partitioning strategy achieves a significant constant-level reduction in computing power, alleviating the computational bottleneck of multi-scale long-range perception in time series modeling. Relying on the gradient flow during model training, the network can spontaneously drive $\Delta p_{n}$ to approximate the deviated true non-integer period during the standard gradient descent optimization process, thereby gradually alleviating the quantization bias introduced by the discrete Fourier transform in the continuous spatial domain. This implicitly mitigates the picket-fence effect without significantly increasing the computational overhead.
		
		\subsection{Asymmetric Routing and Multi-scale Cascade Architecture}
		\begin{figure}[htbp]
			\centering
			\includegraphics[width=0.92\linewidth]{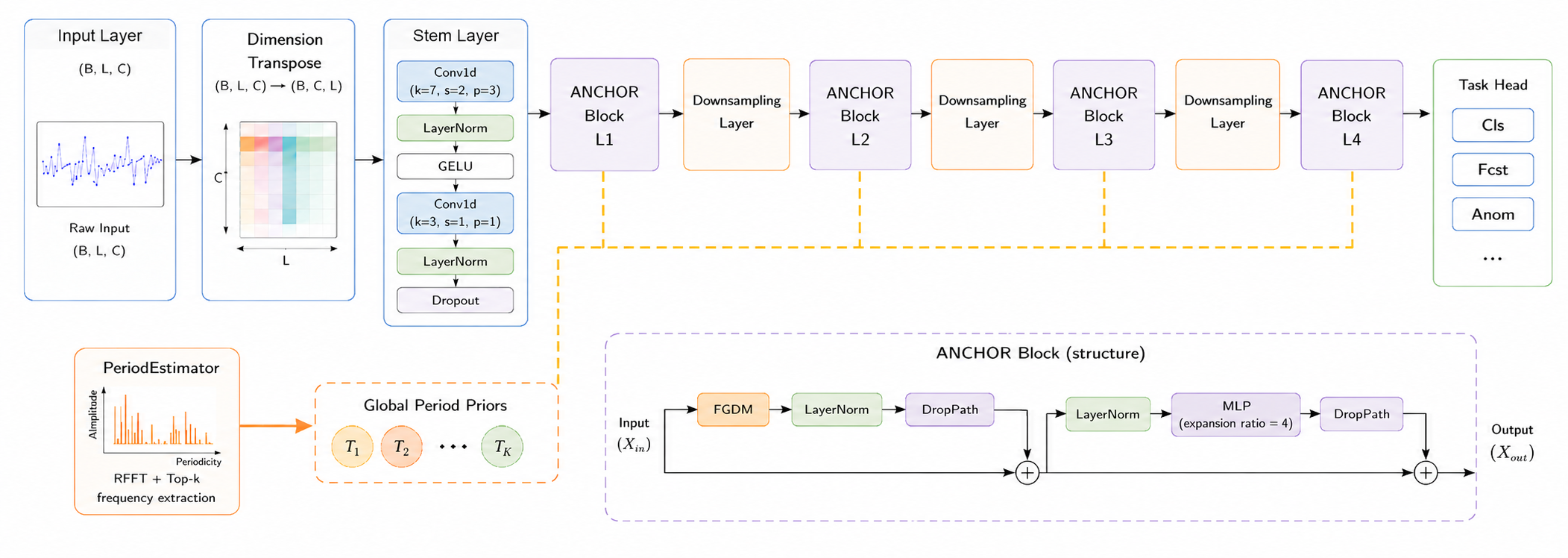}
			\caption{Overall architecture of the ANCHOR model}
			\label{fig:anchor}
		\end{figure}
		
		Inspired by the topological structure of advanced vision meta-architectures such as InternImage, we designed ANCHOR. The core of ANCHOR is the Frequency-Guided Deformable Module. It implements an asymmetric routing strategy by orthogonally partitioning input channels into multiple subspaces, and performs progressive processing across cascaded stages. Each stage utilizes a deformable operator with varying kernel sizes, whose dilation rate is dynamically updated by the macroscopic period prior extracted by the RFFT. This design enables the model to simultaneously preserve high-frequency strong signals and integrate low-frequency weak features. The overall backbone network of ANCHOR is a hierarchical cascade architecture. It includes a stem layer and multiple downsampling stages, with each stage consisting of stacked FGDM blocks. The global period prior from the RFFT extractor is synchronously distributed to each stage, guiding the dilation rate of the DefOp to achieve continuous phase alignment. This architecture effectively models the multi-scale nested characteristics of time series from macroscopic to microscopic levels, as illustrated in Fig.~\ref{fig:anchor}.
		
		\section{Experiment}
		\begin{table}[htbp]
			\centering
			\caption{Summary of experiment benchmarks}
			\label{tab:benchmarks}
			\small
			\begin{tabular}{llll}
				\toprule
				Tasks & Benchmarks & Metrics & Series Length \\
				\midrule
				Forecasting & Short-term: M4 (6 subsets) & SMAPE, MASE, OWA & 6--48 \\
				Classification & UEA (18 subsets) & Accuracy & 6--1751 \\
				\midrule
				\multirow{2}{*}{Anomaly Detection} & SMD, SWaT, PSM & Precision, Recall, $F1$-score & 100 \\
				\cmidrule{2-4} 
				& UCR (203 subsets) & $F1, F1_e, F1_d$, AUPRC & 96 \\
				\bottomrule
			\end{tabular}
		\end{table}
		To validate the universality of ANCHOR, this study conducts comprehensive experiments across three mainstream time series analysis tasks: short-term forecasting, sequence classification, and anomaly detection. All models are uniformly executed in an NVIDIA A100 GPU environment. The short-term forecasting task strictly adheres to the evaluation standards of the M4 dataset. In the sequence classification task, the maximum training epoch is set to 100, with an early stopping patience of 10. The anomaly detection task adopts an evaluation paradigm based on reconstruction error, with a batch size of 128, training epochs of 10, and an anomaly proportion threshold following the experimental settings of TimesNet. For anomaly detection tasks based on the UCR dataset, a forecasting paradigm is employed. To preserve the original dynamic characteristics of the time series, differencing operations and additional preprocessing strategies are not applied; instead, a sliding window of length 96 is directly used for sequence segmentation. Detailed parameter configurations for each experimental baseline are presented in Table~\ref{tab:benchmarks}.

		\subsection{Short-term Forecasting}
		
		The performance evaluation of the short-term forecasting task is conducted based on the M4 benchmark dataset \cite{makridakis2018m4}. This dataset contains 100,000 independent time series, covering multi-domain data sources and multi-scale sampling frequencies. Such highly heterogeneous temporal characteristics introduce significant distribution shifts and frequency variations, constituting the core intervention variables for testing the fitting accuracy and generalization robustness of the models.
		\begin{figure}[htbp]
			\centering
			\includegraphics[width=0.85\linewidth]{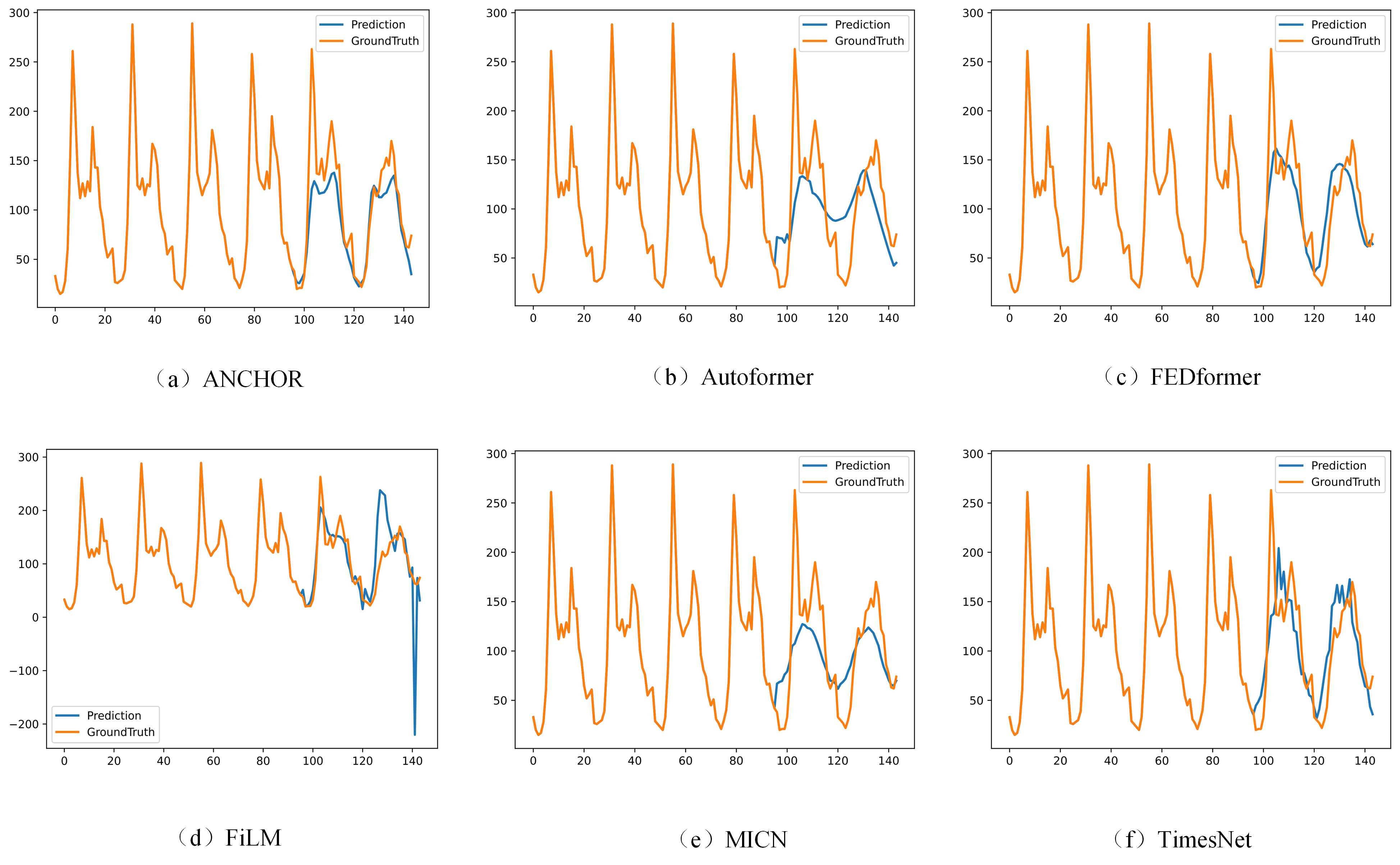}
			\caption{Visualization of M4-Hourly short-term forecasting results given by models}
			\label{fig:visualization}
		\end{figure} 
		
		The quantitative comparison results of various models under the short-term forecasting task are detailed in Table~\ref{tab:m4}. Under the condition of controlling the aforementioned highly heterogeneous variables, the ANCHOR model achieves the optimal performance in most sub-tasks regarding SMAPE, MASE, and the core comprehensive metric OWA. Mechanistic analysis indicates that compared to existing mainstream baseline models based on MLP and Transformer architectures, as well as cross-modal pre-training paradigms introducing LLM/VLM, ANCHOR demonstrates superior representation capabilities when dealing with distribution shifts and multi-scale frequency interference caused by multi-source data. To provide a direct visual validation of our model, we compare its outputs against those of existing baseline models. The detailed qualitative visualizations are presented in Fig.~\ref{fig:visualization}. 
		\begin{table}[H]
			\centering
			\caption{Short-term forecasting task on M4. Bold values indicate the best performance, underlined values denote the second-best results.}
			\label{tab:m4}
			
			\begin{subtable}{\textwidth}
				\centering
				\caption*{(a)}
				\label{tab:m4_part1}
				\scriptsize
				\setlength{\tabcolsep}{4pt}
				\begin{tabular}{llccccc}
					\toprule
					Datasets & Metrics & ANCHOR(Proposed) & SE-LLM(2026) \cite{liu2025semantic} & TimeMixer++(2025) \cite{wang2024timemixer++} & FSCA(2025) \cite{hu2025context} & Time-VLM(2025) \cite{zhong2025time} \\
					\midrule
					Yearly & SMAPE & \textbf{13.032} & 13.294 & 13.397 & 13.288 & \underline{13.285} \\
					& MASE  & \textbf{2.917}  & \underline{2.970}  & 2.990  & 2.974  & 2.993  \\
					& OWA   & \textbf{0.766}  & \underline{0.780}  & 0.786  & 0.781  & 0.783  \\
					\midrule
					Quarterly & SMAPE & \textbf{10.002} & 10.079 & 10.206 & 10.037 & 10.218 \\
					& MASE  & \textbf{1.151}  & 1.177  & 1.201  & 1.174  & 1.203  \\
					& OWA   & \textbf{0.875}  & 0.887  & 0.901  & 0.884  & 0.903  \\
					\midrule
					Monthly & SMAPE & \textbf{12.591} & \underline{12.618} & 12.720 & 12.762 & 12.788 \\
					& MASE  & 0.943 & \textbf{0.931} & 0.943 & 0.947 & 0.942 \\
					& OWA   & \underline{0.880} & \textbf{0.875} & 0.884 & 0.888 & 0.886 \\
					\midrule
					Others & SMAPE & \textbf{4.458} & 4.896 & \underline{4.593} & 4.761 & 4.945 \\
					& MASE  & \textbf{3.065} & 3.306 & 3.380 & 3.207 & 3.257 \\
					& OWA   & \textbf{0.952} & 1.036 & 1.016 & 1.007 & 1.034 \\
					\midrule
					Average & SMAPE & \textbf{11.664} & \underline{11.778} & 11.866 & 11.829 & 11.894 \\
					& MASE  & \textbf{1.553} & \underline{1.578} & 1.598 & 1.581 & 1.592 \\
					& OWA   & \textbf{0.836} & \underline{0.847} & 0.855 & 0.849 & 0.855 \\
					\bottomrule
				\end{tabular}
			\end{subtable}
			
			\vspace{0.3cm}
			
			\begin{subtable}{\textwidth}
				\centering
				\caption*{(b)}
				\label{tab:m4_part2}
				\scriptsize
				\setlength{\tabcolsep}{4pt}
				\begin{tabular}{llccccc}
					\toprule
					Datasets & Metrics & AutoTimes(2024) \cite{liu2024autotimes} & S$^2$IP-LLM(2024) \cite{pan2024textbf} & Time-LLM(2024) \cite{jin2023time} & FPT(2024) \cite{zhou2023one} & iTransformer(2024) \cite{liu2023itransformer} \\
					\midrule
					Yearly & SMAPE & 13.319 & 13.413 & 13.419 & 13.531 & 13.652 \\
					& MASE  & 3.021  & 3.024  & 3.005  & 3.015  & 3.095 \\
					& OWA   & 0.788  & 0.791  & 0.789  & 0.793  & 0.807 \\
					\midrule
					Quarterly & SMAPE & \underline{10.020} & 10.352 & 10.110 & 10.177 & 10.353 \\
					& MASE  & \underline{1.162}  & 1.228  & 1.178  & 1.194  & 1.209 \\
					& OWA   & \underline{0.879}  & 0.918  & 0.889  & 0.898  & 0.911 \\
					\midrule
					Monthly & SMAPE & 12.696 & 12.995 & 12.980 & 12.894 & 13.079 \\
					& MASE  & \underline{0.936} & 0.970 & 0.963 & 0.956 & 0.974 \\
					& OWA   & \underline{0.880} & 0.907 & 0.903 & 0.897 & 0.911 \\
					\midrule
					Others & SMAPE & 4.916 & 4.805 & 4.795 & 4.940 & 4.780 \\
					& MASE  & 3.310 & 3.247 & \underline{3.178} & 3.228 & 3.231 \\
					& OWA   & 1.039 & 1.017 & \underline{1.006} & 1.029 & 1.012 \\
					\midrule
					Average & SMAPE & 11.808 & 12.047 & 11.983 & 11.991 & 12.142 \\
					& MASE  & 1.588 & 1.618 & 1.595 & 1.600 & 1.631 \\
					& OWA   & 0.851 & 0.867 & 0.859 & 0.860 & 0.874 \\
					\bottomrule
				\end{tabular}
			\end{subtable}
		\end{table}
		
		\subsection{Anomaly Detection}
		
		Given the extremely high cost of point-level annotation for rare anomaly events in large-scale industrial monitoring data, this study focuses on the unsupervised time series anomaly detection task, aiming to precisely locate anomalous time nodes within the sequences. Model evaluation is conducted based on three widely used standard benchmark datasets: SMD \cite{su2019robust}, SWaT \cite{mathur2016swat}, and PSM \cite{abdulaal2021practical}. In the data preprocessing stage, this study follows the sliding window mechanism of Anomaly Transformer \cite{xu2021anomaly} to partition continuous sequences into non-overlapping independent segments. To ensure fairness, the experiments uniformly adopt the classic reconstruction task in unsupervised learning as the evaluation framework, utilizing reconstruction error as the anomaly criterion, as shown in Table~\ref{tab:anomaly}. The asterisk denotes that the original Anomaly Transformer employs Temporal Association and Reconstruction Error as joint criteria. To ensure fair comparison, we align the evaluation metric with each model's foundational paradigm: utilizing native prediction error for forecasting-based models (e.g., KAN-AD), while isolating the fundamental reconstruction error for reconstruction-based architectures, deliberately excluding composite criteria.
		
		\begin{table}[H]
			\centering
			\caption{Anomaly detection task comparison}
			\label{tab:anomaly}
			\small
			\setlength{\tabcolsep}{5pt}
			\renewcommand{\arraystretch}{1.15}
			\resizebox{\textwidth}{!}{
				\begin{tabular}{lccc|ccc|ccc}
					\toprule
					\multirow{2}{*}{Model} & \multicolumn{3}{c|}{SMD} & \multicolumn{3}{c|}{SWaT} & \multicolumn{3}{c}{PSM} \\
					& Precision & Recall & F1-score & Precision & Recall & F1-score & Precision & Recall & F1-score \\
					\midrule
					ANCHOR (Proposed) & 86.23 & 81.37 & 83.73 & 94.32 & 95.03 & \textbf{94.68} & 99.21 & 96.54 & \textbf{97.86} \\
					KAN-AD (2025) \cite{zhou2024kan} & / & / & 84.29 & / & / & 93.50 & / & / & 96.50 \\
					UniTS-ST (2025) \cite{gao2024units} & 89.32 & 86.90 & \textbf{88.09} & 92.37 & 94.17 & 93.26 & 98.62 & 96.28 & \underline{97.44} \\
					ModernTCN (2024) \cite{luo2024moderntcn} & 87.86 & 83.85 & \underline{85.81} & 91.83 & 95.98 & \underline{93.86} & 98.09 & 96.38 & 97.23 \\
					DLinear (2023) \cite{zeng2023transformers} & 83.62 & 71.52 & 77.10 & 80.91 & 95.30 & 87.52 & 98.28 & 89.26 & 93.55 \\
					Anomaly Transformer* (2022) \cite{xu2021anomaly} & 88.91 & 82.23 & 85.44 & 72.51 & 97.32 & 83.10 & 68.35 & 94.72 & 79.40 \\
					Non-Stationary Transformers (2022) \cite{liu2022non} & 88.33 & 81.21 & 84.62 & 68.03 & 96.75 & 79.89 & 97.82 & 96.76 & 97.29 \\
					Informer (2021) \cite{zhou2021informer} & 86.60 & 77.23 & 81.65 & 70.29 & 96.75 & 81.43 & 64.27 & 96.33 & 77.10 \\
					\bottomrule
			\end{tabular}}
		\end{table}
		\begin{table}[htbp]
			\centering
			\caption{Performance Comparison on UCR Archive}
			\label{tab:anomaly_UCR}
			\small 
			\begin{tabular}{l cccc}
				\toprule
				Method & {F1} & {$F1_e$ (Event)} & {$F1_d$ (Delay)} & {AUPRC} \\
				\midrule
				SRCNN (2019) \cite{ren2019time}                                         & 0.5964 & 0.1369 & 0.1656 & 0.5109 \\
				SAND (2021) \cite{boniol2021sand}              & 0.7044 & {\underline{0.5108}} & {\underline{0.5116}} & 0.6550 \\
				Anomaly Transformer (2022) \cite{xu2021anomaly}& 0.6135 & 0.1696 & 0.1084 & 0.5458 \\
				TranAD (2022) \cite{tuli2022tranad}            & 0.5278 & 0.1840 & 0.1554 & 0.4599 \\
				SubLOF (2000) \cite{breunig2000lof}            & {\textbf{0.8468}} & 0.4772 & 0.4151 & {\underline{0.8001}} \\
				TimesNet (2023) \cite{wu2022timesnet}          & 0.5273 & 0.1805 & 0.1439 & 0.4536 \\
				FITS (2023) \cite{xu2023fits}                  & 0.6664 & 0.2926 & 0.2912 & 0.5969 \\
				OFA (2023) \cite{zhou2023one}                  & 0.6294 & 0.3176 & 0.1503 & 0.5699 \\
				FCVAE (2024) \cite{wang2024revisiting}         & 0.7651 & 0.3812 & 0.2857 & 0.7145 \\
				LSTMAD (2015) \cite{malhotra2015long}          & 0.7040 & 0.3482 & 0.3121 & 0.6432 \\
				KAN (2024) \cite{liu2024kan}                   & {\underline{0.8016}} & 0.4120 & 0.3971 & 0.7489 \\
				\midrule
				ANCHOR (Proposed)                     & 0.7890 & {\textbf{0.5243}} & {\textbf{0.6259}} & {\textbf{0.8112}} \\
				\bottomrule
			\end{tabular}
		\end{table}
		Under the unified unsupervised reconstruction error framework, the performance of the ANCHOR model in multivariate time series anomaly detection surpasses most existing mainstream architectures. Experiments demonstrate that, despite occasional local oscillations in extreme non-stationary scenarios, ANCHOR maintains a high degree of robustness when processing large-scale high-dimensional temporal features.
		
		A systematic evaluation of ANCHOR is conducted on the UCR \cite{wu2021current} univariate time series repository. The experiments follow the prediction paradigm across 203 subsets of the UCR archive. Specifically, the model utilizes a sliding window of length 96 to extract local temporal features and performs point-wise prediction. To ensure experimental fairness, a unified data splitting, training pipeline, and evaluation protocol are applied across all datasets, with independent models trained for each individual sequence. The comparative analysis includes 10 representative baseline models, encompassing both traditional methods and deep learning architectures. All baselines adhere strictly to the implementation details and hyperparameter configurations reported in their original publications. In instances where specific configurations were not provided, a unified grid search strategy is employed for optimization to minimize the impact of implementation variance on the comparative results.
		
		Considering that industrial operations and maintenance prioritize the identification quality of anomalous events and the timeliness of alarms, this study eschews the Point Adjustment (PA) strategy, which is prone to introducing evaluation bias. Instead, a comprehensive evaluation is conducted using Event F1, Delay F1, F1, and AUPRC. Specifically, Event F1 measures the coverage effectiveness of anomalous intervals at the event granularity, while Delay F1 characterizes the model's responsiveness during the onset of an anomaly. Experimental results demonstrate that ANCHOR achieves the best performance across Event F1, Delay F1, and AUPRC, indicating its superiority in anomalous event coverage, early detection timeliness, and overall ranking capability, as shown in Table~\ref{tab:anomaly_UCR}. Although certain baseline methods exhibit higher point-level F1 scores, their event-level detection capabilities and delay responsiveness remain relatively limited, suggesting that a single point-level metric is insufficient to reflect a model's actual utility in practical applications. Overall, ANCHOR demonstrates superior comprehensive performance in balancing detection effectiveness with early warning timeliness.
		
		Time series classification tasks possess core evaluative value in fields such as pattern recognition and clinical auxiliary diagnosis \cite{goldberger2000components}.To validate the model's capability in extracting high-order global semantic representations, this study relies on the UEA multivariate time series classification archive \cite{bagnall2018uea}, introducing sequence-level classification tasks as the core evaluation benchmark. Regarding benchmark construction, to ensure the objectivity of the evaluation protocol and its alignment with the model's inductive biases, this study predefined the following dataset screening criteria: inputs must be raw time-domain continuously sampled sequences rather than pre-extracted frequency-domain features; sequences must exhibit a certain degree of temporal continuity, local smoothness, or spectral decomposability; the data preprocessing procedure must not introduce large-scale zero-padding or excessive discretization artifacts; and the sequence length must fall within the permissible range of current experimental resources. Based on the aforementioned criteria, 18 datasets were selected from the UEA archive as the primary evaluation set.These datasets encompass scenarios such as gesture and motion capture, speech signal mapping, and physiological and medical monitoring. Concurrently, during the data preprocessing stage, this study strictly adheres to the standardization protocol established by Zerveas et al. \cite{zerveas2021transformer}, preserving the heterogeneity of sequence lengths across the selected subsets to objectively examine the model's capacity to process variable-length inputs. As evidenced by the quantitative results in Table~\ref{tab:classification}, ANCHOR effectively captures the discriminative features across both macroscopic trends and microscopic fluctuations. This comprehensive feature extraction allows ANCHOR to establish state-of-the-art benchmarks in scenarios requiring high-fidelity global representations, while maintaining robust and competitive accuracy across diverse multi-source monitoring datasets.
		
		\begin{table}[H]
			\centering
			\caption{Model comparison in classification. Bold values indicate the best performance, underlined values denote the second-best results.}
			\label{tab:classification}
			
			\begin{subtable}{\textwidth}
				\centering
				\caption*{(a)}
				\label{tab:classification_part1}
				\scriptsize
				\setlength{\tabcolsep}{4pt}
				\begin{tabular}{lcccc}
					\toprule
					Datasets & ANCHOR (Proposed) & MambaSL(2026) \cite{jungmambasl} & TSC Mamba(2025) \cite{ahamed2025tscmamba} & InterpGN(2025) \cite{wen2025shedding} \\
					\midrule
					EthanolConcentration & \underline{38.412} & \textbf{42.586} & 30.798 & 28.897 \\
					FaceDetection & \textbf{71.216} & 69.296 & \underline{70.204} & 63.791 \\
					Handwriting & 33.562 & \underline{60.824} & 45.059 & 58.706 \\
					Heartbeat & 78.612 & \underline{80.488} & 78.537 & \underline{80.488} \\
					JapaneseVowels & \textbf{99.912} & 98.649 & 98.378 & \underline{99.730} \\
					PEMS-SF & \underline{89.342} & 85.549 & \textbf{90.173} & 86.127 \\
					SelfRegulationSCP1 & \textbf{93.951} & 92.491 & 92.833 & 90.444 \\
					SelfRegulationSCP2 & 55.432 & \textbf{65.000} & 62.222 & 58.333 \\
					SpokenArabicDigits & \textbf{99.963} & \underline{99.955} & 95.816 & 99.864 \\
					UWaveGestureLibrary & 88.623 & 93.438 & \textbf{95.313} & 88.438 \\
					ArticularyWordRecognition & \textbf{99.667} & \underline{99.333} & \underline{99.333} & 99.000 \\
					BasicMotions & \textbf{100.000} & \textbf{100.000} & \textbf{100.000} & \textbf{100.000} \\
					CharacterTrajectories & \textbf{99.867} & 99.721 & 99.513 & \underline{99.791} \\
					Cricket & 98.231 & \textbf{100.000} & \textbf{100.000} & \textbf{100.000} \\
					FingerMovements & 67.000 & \underline{71.000} & 66.000 & 67.000 \\
					HandMovementDirection & \underline{72.662} & 70.270 & 60.811 & 45.946 \\
					NATOPS & 98.121 & \underline{98.889} & 92.222 & \underline{98.889} \\
					RacketSports & 88.568 & \underline{92.763} & 90.789 & 91.447 \\
					\bottomrule
				\end{tabular}
			\end{subtable}
			
			\vspace{0.3cm}
			
			\begin{subtable}{\textwidth}
				\centering
				\caption*{(b)}
				\label{tab:classification_part2}
				\scriptsize
				\setlength{\tabcolsep}{4pt}
				\begin{tabular}{lcccc}
					\toprule
					Datasets & Time Mixer++(2025) \cite{wang2024timemixer++} & TSLANet(2024) \cite{eldele2024tslanet} & Modern TCN(2024) \cite{luo2024moderntcn} & TimesNet(2023) \cite{wu2022timesnet} \\
					\midrule
					EthanolConcentration & 34.601 & 31.939 & 32.319 & 33.460 \\
					FaceDetection & 69.665 & 66.969 & 66.998 & 70.148 \\
					Handwriting & 33.647 & \textbf{62.000} & 30.118 & 37.529 \\
					Heartbeat & \underline{80.488} & 79.512 & 78.049 & \textbf{83.902} \\
					JapaneseVowels & 99.189 & 98.919 & 98.649 & 98.649 \\
					PEMS-SF & 89.017 & 89.017 & 86.705 & 87.861 \\
					SelfRegulationSCP1 & 92.150 & 89.761 & \underline{93.515} & 92.150 \\
					SelfRegulationSCP2 & 60.556 & \underline{63.889} & 60.000 & 60.000 \\
					SpokenArabicDigits & 99.545 & 98.045 & 99.500 & 99.545 \\
					UWaveGestureLibrary & 90.625 & \underline{94.375} & 88.438 & 88.438 \\
					ArticularyWordRecognition & 99.000 & \underline{99.333} & 98.667 & 98.667 \\
					BasicMotions & \textbf{100.000} & \textbf{100.000} & \textbf{100.000} & \textbf{100.000} \\
					CharacterTrajectories & 99.373 & 99.164 & 99.304 & 98.747 \\
					Cricket & \underline{98.611} & 91.667 & 97.222 & 97.222 \\
					FingerMovements & 67.000 & 68.000 & 67.000 & 66.000 \\
					HandMovementDirection & \textbf{72.973} & 48.649 & 66.216 & \textbf{72.973} \\
					NATOPS & 96.111 & \textbf{99.444} & 95.556 & 98.333 \\
					RacketSports & 91.447 & \textbf{94.079} & 84.211 & 89.474 \\
					\bottomrule
				\end{tabular}
			\end{subtable}
			
			\vspace{0.3cm}
			
			\begin{subtable}{\textwidth}
				\centering
				\caption*{(c)}
				\label{tab:classification_part3}
				\scriptsize
				\setlength{\tabcolsep}{4pt}
				\begin{tabular}{lccc}
					\toprule
					Datasets & DLinear(2023) \cite{zeng2023transformers} & FEDformer(2022) \cite{zhou2022fedformer} & DTW(1994) \\
					\midrule
					EthanolConcentration & 31.179 & 33.840 & 32.319 \\
					FaceDetection & 69.495 & 69.892 & 52.866 \\
					Handwriting & 23.765 & 33.412 & 60.706 \\
					Heartbeat & 76.585 & 78.049 & 71.707 \\
					JapaneseVowels & 96.486 & 97.568 & 95.946 \\
					PEMS-SF & 82.659 & 89.017 & 71.098 \\
					SelfRegulationSCP1 & 92.833 & 75.085 & 77.474 \\
					SelfRegulationSCP2 & 57.222 & 59.444 & 53.889 \\
					SpokenArabicDigits & 96.953 & 99.454 & 97.226 \\
					UWaveGestureLibrary & 82.813 & 79.688 & 90.313 \\
					ArticularyWordRecognition & 97.667 & 92.333 & 98.667 \\
					BasicMotions & 87.500 & 95.000 & \underline{97.500} \\
					CharacterTrajectories & 98.120 & 97.284 & 98.816 \\
					Cricket & 91.667 & 90.278 & \textbf{100.000} \\
					FingerMovements & 62.000 & \textbf{72.000} & 53.000 \\
					HandMovementDirection & 66.216 & 71.622 & 18.919 \\
					NATOPS & 96.111 & 96.111 & 88.333 \\
					RacketSports & 78.947 & 84.868 & 80.263 \\
					\bottomrule
				\end{tabular}
			\end{subtable}
		\end{table}
		\subsection{Ablation Study}
		
		\subsubsection{Ablation Study on the Temporal Representation Mechanism of ANCHOR}
		
		To further dissect the specific contributions of the core designs of ANCHOR to model performance, this paper constructed three progressive sets of ablation experiments centered around the deformable offset mechanism and the Gaussian radial basis function. Regarding the experimental setup, all variants maintained strict consistency in the number of network layers, channel dimensions, and hyperparameters to adhere to the principle of controlling a single variable. First, ANCHOR-1D employing standard 1D convolution is used as the baseline. Second, ANCHOR-BL introduces a 1D-adapted DCNv4 module with bilinear interpolation. Finally, the complete model ANCHOR replaces the interpolation kernel with a Gaussian RBF. To verify the generalization capability of the aforementioned components under different data distributions, this study selected three major categories of representative benchmarks: the short-term forecasting tasks M4 Yearly and Others, which contain macroscopic and microscopic periodic trends; the anomaly detection tasks SMD and MSL, which feature multidimensional sudden anomalies; and the variable-length and high-noise time series classification tasks JapaneseVowels and SelfRegulationSCP1 from the UEA archive. Quantitative experimental results display specific evaluation metrics such as MSE, MAE, and Accuracy, which are detailed in Table~\ref{tab:ablation1}, where JV and SRS1 denote the JapaneseVowels and SelfRegulationSCP1 datasets, respectively.
		
		\begin{table}[H]
			\centering
			\caption{Ablation study on the temporal representation mechanism of ANCHOR}
			\label{tab:ablation1}
			\small
			\renewcommand{\arraystretch}{1.15}
			\begin{tabular}{llllll}
				\toprule
				Task & Dataset & Metric & ANCHOR-1D & ANCHOR-BL & ANCHOR-Gaussian \\
				\midrule
				Short-term Forecasting & M4-Yearly & SMAPE & 13.289 & 13.157 & \textbf{13.032} \\
				Short-term Forecasting & M4-Yearly & MASE & 3.012 & 2.963 & \textbf{2.917} \\
				Short-term Forecasting & M4-Yearly & OWA & 0.785 & 0.775 & \textbf{0.766} \\
				Short-term Forecasting & M4-Others & SMAPE & 5.022 & 4.612 & \textbf{4.458} \\
				Short-term Forecasting & M4-Others & MASE & 3.201 & 3.123 & \textbf{3.065} \\
				Short-term Forecasting & M4-Others & OWA & 1.033 & 0.978 & \textbf{0.952} \\
				Anomaly Detection & SMD & F1-score & 79.16 & 82.56 & \textbf{83.73} \\
				Anomaly Detection & PSM & F1-score & 96.25 & 97.75 & \textbf{97.86} \\
				Classification & JV & Accuracy & 96.81 & 98.92 & \textbf{99.91} \\
				Classification & SRS1 & Accuracy & 89.76 & 93.94 & \textbf{93.95} \\
				\bottomrule
			\end{tabular}
		\end{table}
		
		The overall performance of ANCHOR-BL across multiple tasks is significantly superior to that of ANCHOR-1D, confirming the advantage of adaptive sampling. Further comparison reveals that the complete model obtains a secondary gain after introducing the Gaussian RBF, indicating that the smooth and continuous derivative characteristics of Gaussian interpolation improve optimization stability and local feature coherence.
		
		\subsubsection{Comparative Experiment on Quantization Fragment Compensation Capability}
		
\begin{figure}[htbp]
			\centering
			\includegraphics[width=0.85\linewidth]{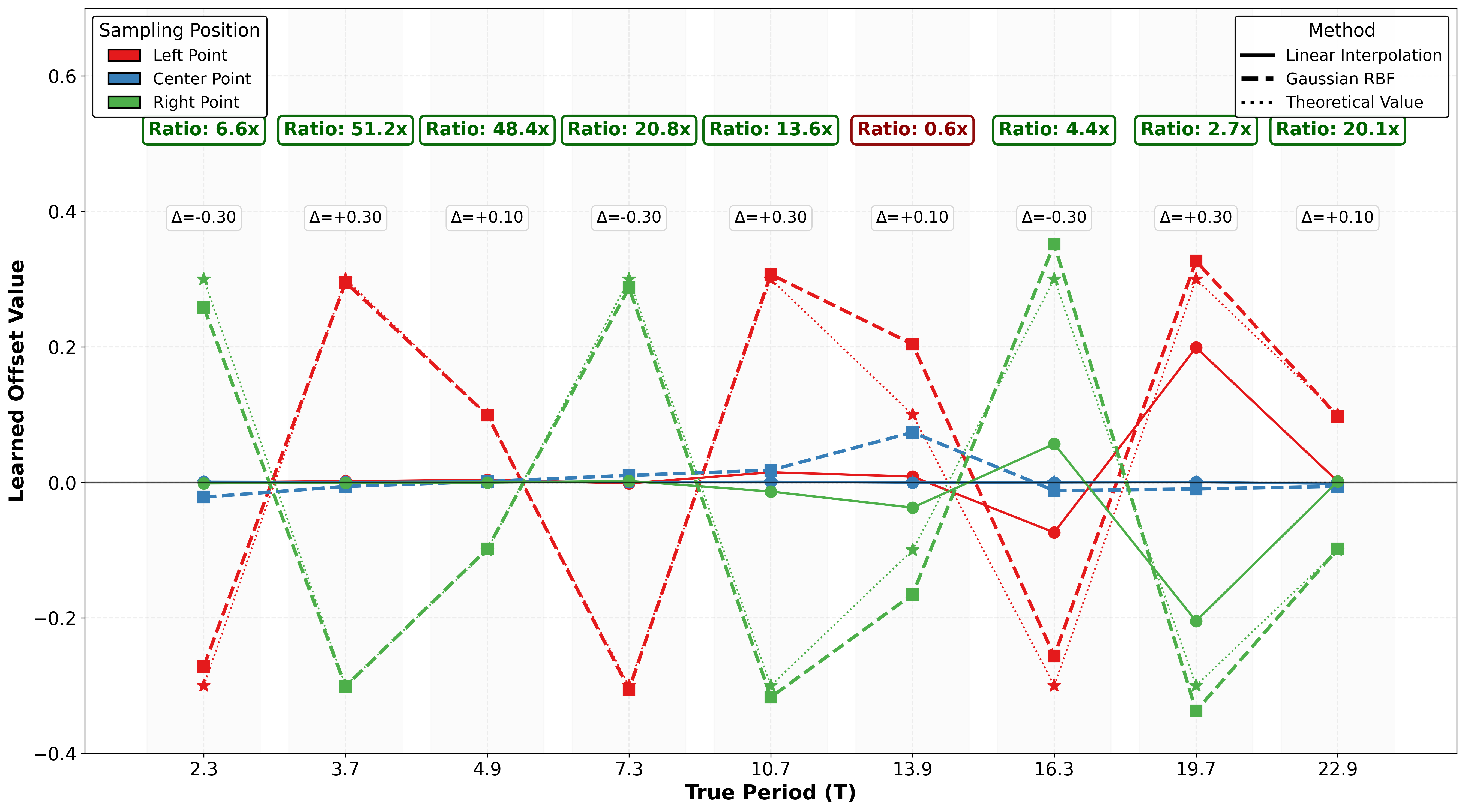}
			\caption{Comparative results of quantization fragment compensation capability}
			\label{fig:compensation}
		\end{figure}

		To quantitatively and intuitively validate the advantages of the one-dimensional Gaussian radial basis function operator in microscopic phase alignment, controlled comparative experiments were designed to target signals with non-integer periods. When the model can only rely on discrete dilation rates as priors, the true physical phase of the signal often cannot perfectly coincide with the sampling positions, thereby generating sub-pixel level positional errors. Although such errors are minute, they directly impact the precision of phase alignment. To this end, 9 distinct fractional periods were designed, ensuring that the theoretical offset for each period is non-zero. For each period, the dynamic integer period was computed and subsequently trained two models utilizing linear interpolation and Gaussian interpolation, respectively, to learn the quantization fragment compensation. The results are illustrated in Fig.~\ref{fig:compensation}.
		
		The mathematical definition of the error ratio is as follows:
		\[
		\eta = \frac{\mathcal{E}_{linear}}{\mathcal{E}_{gaussian}},
		\]
		Where $\eta$ represents the error ratio, utilized to quantify the performance gain amplitude of the Gaussian interpolator relative to the baseline. $\eta > 1$ indicates that Gaussian interpolation holds an advantage. $\mathcal{E}_{linear}$ denotes the prediction error when adopting traditional linear interpolation. $\mathcal{E}_{gaussian}$ denotes the prediction error when utilizing Gaussian radial basis interpolation. During the training process, MSE is employed as the loss function to optimize the model parameters. When evaluating the experimental results, the MAE between the learned offset and the theoretical offset is calculated, in order to more intuitively reflect the precision discrepancy between the two interpolation methods. Experimental results demonstrate that traditional linear interpolation, due to its inherently piecewise linear nature, provides insufficiently continuous gradients near certain positions and may even suffer from a lack of gradient information. Consequently, when optimizing minute phase offsets, the model is prone to halting prematurely near suboptimal solutions, making it difficult to further approximate the true alignment position. In contrast, the Gaussian RBF operator possesses a smoother response and more continuous gradient characteristics, enabling it to provide stable and effective optimization signals at the majority of non-integer coordinates.
		
		\subsubsection{Sensitivity Analysis of Top-k Frequency Components}
\begin{figure}[htbp]
			\centering
			\includegraphics[width=0.85\linewidth]{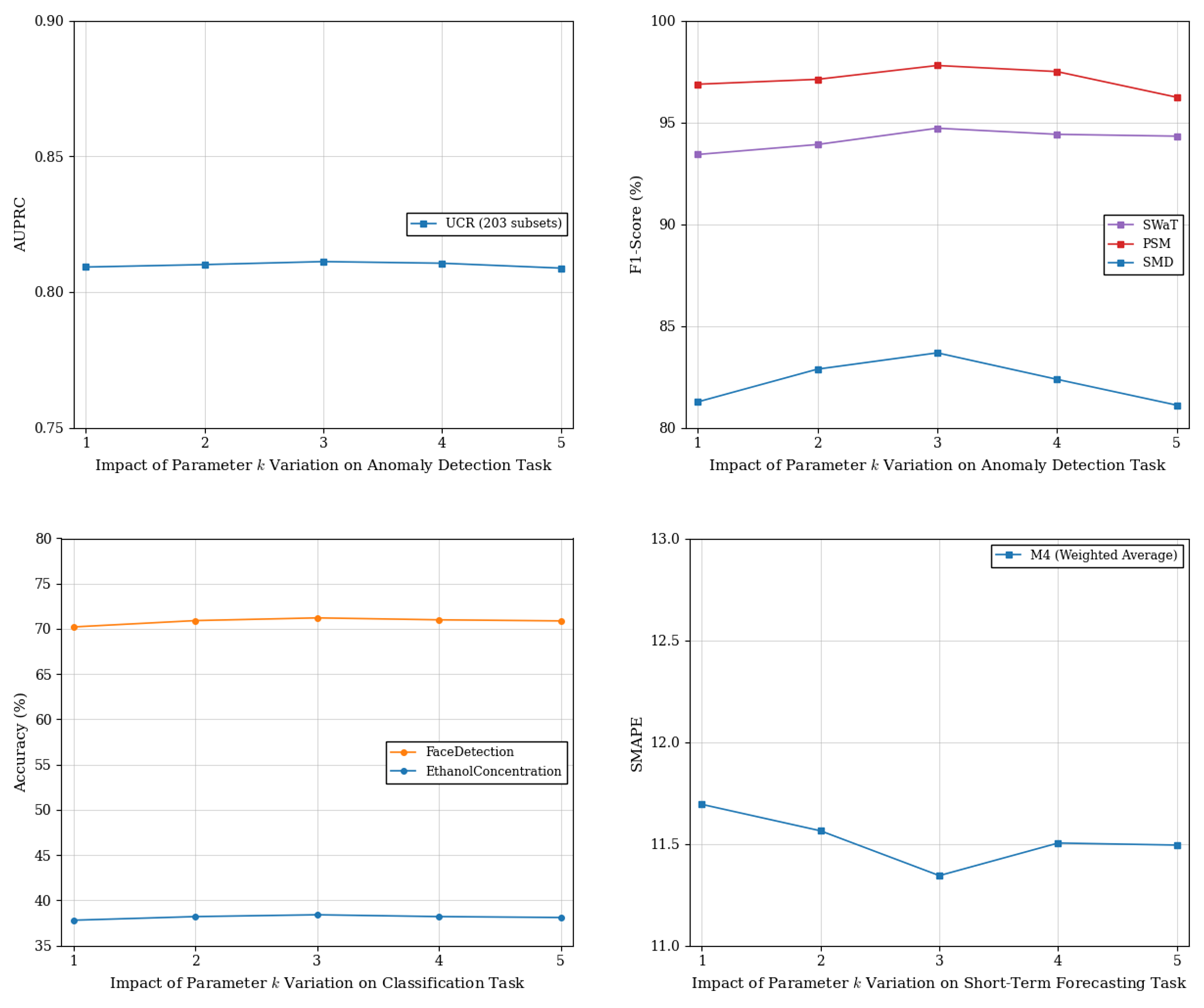}
			\caption{Sensitivity Analysis of Top-k Frequency Components}
			\label{fig:top_k}
		\end{figure}

		Regarding the selection of the top-$k$ dominant frequency components, the value of $k$ is determined according to the specific task requirements. Parameter sensitivity experiments demonstrate that the overall model performance is relatively insensitive to variations in $k$, exhibiting marginal fluctuations. However, in tasks that heavily rely on local reconstruction accuracy—such as anomaly detection and short-term forecasting—$k$ exerts subtle yet discernible regularized effects, as shown in Fig.~\ref{fig:top_k}. Comprehensive results indicate that $k=3$ is typically the optimal selection, at which point the SMAPE and F1-score consistently achieve their peak or near-optimal values. An excessively small $k$ leads to the omission of critical periodic information, whereas an excessively large $k$ tends to introduce low-energy high-frequency noise, thereby interfering with time-domain reconstruction and fine-grained detail fitting.
		\subsubsection{Cross-Model Reconstruction Verification of FGDM}
		\begin{figure}[htbp]
			\centering
			\includegraphics[width=0.65\linewidth]{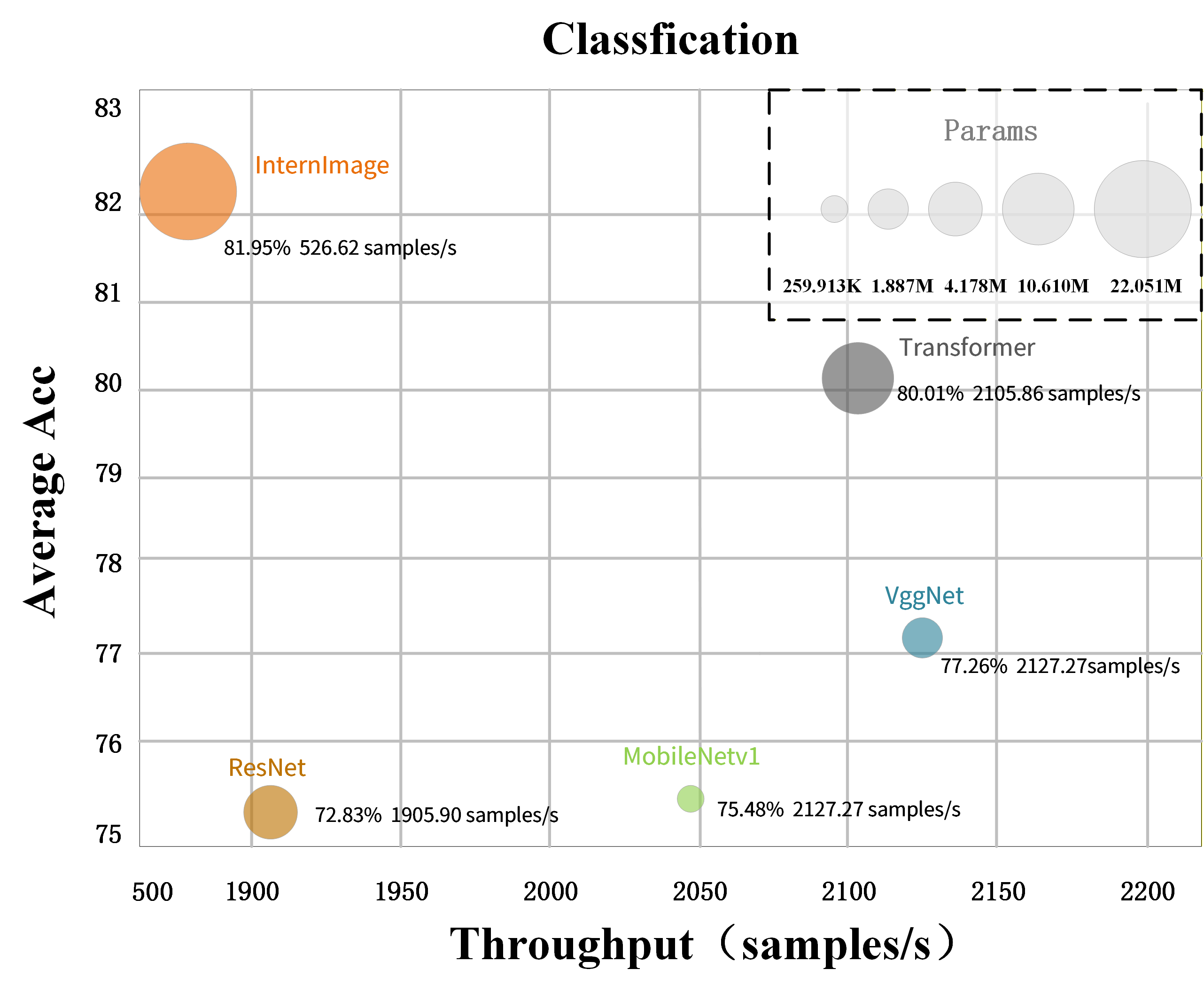}
			\caption{Accuracy-Throughput Trade-off Across Architectures}
			\label{fig:throughput}
		\end{figure}
		To evaluate the structural generalization and deployment compatibility of the FGDM module, it is instantiated into backbone networks spanning diverse design paradigms, including VGGNet, the lightweight MobileNetV1, ResNet, Transformer, and our InternImage-based ANCHOR architecture. Experimental results indicate that FGDM seamlessly integrates into various architectures with stable convergence, validating its high versatility as a cross-architecture plugin for frequency-spatial feature extraction, as shown in Fig.~\ref{fig:throughput}
		
		Among all FGDM-equipped variants, ANCHOR achieved the highest classification accuracy. Although its throughput (526.62 samples/s) is lower than that of lightweight models, this primarily reflects the inherent accuracy-efficiency trade-offs of the respective underlying backbones. This horizontal comparison demonstrates that FGDM is highly adaptable: it can be scaled down to lightweight networks to meet high-throughput edge requirements or deeply coupled with large-scale bases to further maximize performance potential.
		
		While the aforementioned experiments demonstrate FGDM's adaptability within general foundational backbones, we further investigate its scalability and robustness in complex time-series forecasting tasks. To verify its plug-and-play capability within highly customized mainstream forecasters, differentiated reconstructions on three baseline models were performed across the SMD, SWaT, and PSM datasets (Table~\ref{tab:reconstruction}). Specifically, for FEDFormer, while retaining the original encoder-decoder framework and cross-sequence interaction paths, the self-modeling unit was replaced by FGDM to compensate for the architecture's inherent lack of local sensitivity. For Informer, we constructed a dual-stream encoder parallelizing global dependency extraction with an FGDM local branch via adaptive weight fusion, and simultaneously introduced an FGDM enhancement path during the downsampling stage to retain local abrupt changes against compression loss. For TimesNet, the original core module based on 2D period rearrangement was replaced by a 1D local dynamic extraction unit driven by FGDM, while strictly preserving its residual stacking and hierarchical normalization structures. By maintaining the primary frameworks of these baselines, these targeted modifications definitively validate the structural compatibility and universality of FGDM as a cross-architecture local modeling module.
		
		\begin{table}[H]
			\centering
			\caption{Cross-model reconstruction verification of FGDM}
			\label{tab:reconstruction}
			\small
			\setlength{\tabcolsep}{5pt}
			\renewcommand{\arraystretch}{1.15}
			\resizebox{\textwidth}{!}{
				\begin{tabular}{lccc|ccc|ccc}
					\toprule
					\multirow{2}{*}{Models} & \multicolumn{3}{c|}{SMD} & \multicolumn{3}{c|}{SWaT} & \multicolumn{3}{c}{PSM} \\
					& Precision & Recall & F1-score & Precision & Recall & F1-score & Precision & Recall & F1-score \\
					\midrule
					FEDFormer & 72.66 & 80.58 & 76.42 & 99.96 & 65.55 & 79.18 & 99.99 & 81.95 & 90.08 \\
					FEDFormer-FGDM & 72.73 & 81.47 & \textbf{76.85} & 99.98 & 67.51 & \textbf{80.60} & 99.99 & 83.01 & \textbf{90.71} \\
					Informer & 86.60 & 77.23 & 81.65 & 70.29 & 96.75 & 81.42 & 64.27 & 96.33 & 77.10 \\
					Informer-FGDM & 87.93 & 78.67 & \textbf{83.04} & 71.63 & 98.27 & \textbf{82.86} & 66.08 & 97.76 & \textbf{78.86} \\
					TimesNet & 77.95 & 87.71 & 82.54 & 90.50 & 95.34 & 92.86 & 98.41 & 96.11 & 97.25 \\
					TimesNet-FGDM & 78.47 & 88.31 & \textbf{83.10} & 91.02 & 95.82 & \textbf{93.36} & 99.13 & 97.19 & \textbf{98.15} \\
					\bottomrule
			\end{tabular}}
		\end{table}
		
		Building upon the replacement of core operators, the reconstructed models further introduced a period-aware dynamic receptive field adjustment mechanism. By estimating the latent dominant period of the input sequence, this mechanism adaptively adjusts the local attention scale of the FGDM unit, enabling the receptive field to dynamically vary in accordance with the sequence's periodic structure. Experimental results confirm that this systematic reconstruction achieved stable gains across all three datasets, indicating that FGDM can significantly broaden the model's feature capture boundaries for non-stationary local anomalies while preserving the baseline models' long-sequence dependency modeling capabilities.
		
		\subsubsection{Analysis of Frequency-Space Receptive Field Mapping Strategy}
		
		To verify the effectiveness of dynamically allocating receptive fields based on frequency characteristics in feature extraction, this section conducts ablation experiments on three anomaly detection datasets: SMD, SWaT, and PSM. The experiments compare two orders: ANCHOR-Asc adopts a strategy of binding low-frequency components to narrow receptive fields and high-frequency components to wide receptive fields, whereas ANCHOR-Desc adopts the reverse strategy, as shown in Table~\ref{tab:mapping}.
		
		\begin{table}[H]
			\centering
			\caption{Analysis of frequency-space receptive field mapping strategy}
			\label{tab:mapping}
			\small
			\setlength{\tabcolsep}{5pt}
			\renewcommand{\arraystretch}{1.15}
			\resizebox{\textwidth}{!}{
				\begin{tabular}{lccc|ccc|ccc}
					\toprule
					\multirow{2}{*}{Model} & \multicolumn{3}{c|}{SMD} & \multicolumn{3}{c|}{SWaT} & \multicolumn{3}{c}{PSM} \\
					& Precision & Recall & F1-score & Precision & Recall & F1-score & Precision & Recall & F1-score \\
					\midrule
					ANCHOR-Asc & 86.23 & 81.37 & \textbf{83.73} & 94.32 & 95.03 & \textbf{94.68} & 99.21 & 96.54 & \textbf{97.86} \\
					ANCHOR-Desc & 85.11 & 80.24 & 82.60 & 93.04 & 94.12 & 93.58 & 98.23 & 95.54 & 96.87 \\
					\bottomrule
			\end{tabular}}
		\end{table}
		
		The experimental results indicate that ANCHOR-Asc achieves superior performance, primarily because this mechanism aggregates long-range context by allocating wide receptive fields to high-frequency components to suppress local false alarms, while simultaneously focusing on transient gradients by allocating narrow receptive fields to low-frequency components to circumvent the over-smoothing of trends.
		
		\section{Conclusion}
		\label{sec:conclusion}
		This study proposes the ANCHOR architecture, aiming to bridge the structural gap between discrete frequency-domain transformations and continuous spatial phase alignments in time series analysis. By utilizing the RFFT to extract macroscopic explicit periods as physical anchors to navigate microscopic deformation optimization, and introducing an infinitely differentiable 1D Gaussian radial basis operator to remodel the backpropagation error surface, this architecture mathematically alleviates the gradient truncation dilemma inherent in traditional interpolation under sub-pixel offsets, and compensates for the discrete picket-fence effect. Coupled with an asymmetric energy routing mechanism, the model has preliminarily validated the effectiveness of decoupling macroscopic physical trends from microscopic spatial perturbations across multiple benchmark tasks. 
		
		However, it is recognized that this framework retains explicit theoretical limitations in extreme non-stationary scenarios. Its current feature aggregation and interpolation mechanisms remain constrained by empirically fixed hyperparameters, lacking sufficient dynamic perception and adaptive regulation capabilities when confronted with complex industrial noise characterized by strong heterogeneity and extremely low signal-to-noise ratios. To this end, future research will focus on the dynamic reconstruction of physical mechanisms: introducing local time-frequency analysis to achieve dynamic anchor tracking, and upgrading core scale factors into data-driven, adaptive learnable parameters, thereby enhancing representation performance in extreme non-stationary time series.
		
		\bibliographystyle{unsrt}
		\bibliography{references}

\begin{thebibliography}{10}

\bibitem{liang2024foundation}
Yuxuan Liang, Haomin Wen, Yuqi Nie, Yushan Jiang, Ming Jin, Dongjin Song,
  Shirui Pan, and Qingsong Wen.
\newblock Foundation models for time series analysis: A tutorial and survey.
\newblock In {\em Proceedings of the 30th ACM SIGKDD conference on knowledge
  discovery and data mining}, pages 6555--6565, 2024.

\bibitem{qiu2026rethinking}
Xin Qiu, Junlong Tong, Yirong Sun, Yunpu Ma, Wei Zhang, and Xiaoyu Shen.
\newblock Rethinking the role of llms in time series forecasting.
\newblock {\em arXiv preprint arXiv:2602.14744}, 2026.

\bibitem{wang2024timexer}
Yuxuan Wang, Haixu Wu, Jiaxiang Dong, Guo Qin, Haoran Zhang, Yong Liu, Yunzhong
  Qiu, Jianmin Wang, and Mingsheng Long.
\newblock Timexer: Empowering transformers for time series forecasting with
  exogenous variables.
\newblock In A.~Globerson, L.~Mackey, D.~Belgrave, A.~Fan, U.~Paquet,
  J.~Tomczak, and C.~Zhang, editors, {\em Advances in Neural Information
  Processing Systems}, volume~37, pages 469--498. Curran Associates, Inc.,
  2024.

\bibitem{wu2022timesnet}
Haixu Wu, Tengge Hu, Yong Liu, Hang Zhou, Jianmin Wang, and Mingsheng Long.
\newblock Timesnet: Temporal 2d-variation modeling for general time series
  analysis.
\newblock In {\em The Eleventh International Conference on Learning
  Representations}, 2023.

\bibitem{mahesh2025themis}
Yadav Mahesh~Lorik, Kaushik Sarveswaran, Nagaraj Sundaramahalingam, and
  Aravindakumar Venugopalan.
\newblock Themis: Unlocking pretrained knowledge with foundation model
  embeddings for anomaly detection in time series.
\newblock {\em arXiv e-prints}, pages arXiv--2510, 2025.

\bibitem{ahmed2016survey}
Mohiuddin Ahmed, Abdun~Naser Mahmood, and Jiankun Hu.
\newblock A survey of network anomaly detection techniques.
\newblock {\em Journal of network and computer applications}, 60:19--31, 2016.

\bibitem{langer2025opentslm}
Patrick Langer, Thomas Kaar, Max Rosenblattl, Maxwell~A Xu, Winnie Chow, Martin
  Maritsch, Robert Jakob, Ning Wang, Juncheng Liu, Aradhana Verma, et~al.
\newblock Opentslm: Time-series language models for reasoning over multivariate
  medical text-and time-series data.
\newblock {\em arXiv preprint arXiv:2510.02410}, 2025.

\bibitem{goldberger2000components}
Ary~L Goldberger, Lu{\'\i}s Amaral, Leon Glass, Jeffrey~M Hausdorff, Plamen~Ch
  Ivanov, Roger~G Mark, Joseph~E Mietus, George~B Moody, Chung-Kang Peng, and
  H~Eugene Stanley.
\newblock Components of a new research resource for complex physiologic
  signals.
\newblock {\em PhysioBank, PhysioToolkit, and Physionet}, 2000.

\bibitem{gao2024units}
Shanghua Gao, Teddy Koker, Owen Queen, Thomas Hartvigsen, Theodoros
  Tsiligkaridis, and Marinka Zitnik.
\newblock Units: A unified multi-task time series model.
\newblock In A.~Globerson, L.~Mackey, D.~Belgrave, A.~Fan, U.~Paquet,
  J.~Tomczak, and C.~Zhang, editors, {\em Advances in Neural Information
  Processing Systems}, volume~37, pages 140589--140631. Curran Associates,
  Inc., 2024.

\bibitem{box2015time}
George~EP Box, Gwilym~M Jenkins, Gregory~C Reinsel, and Greta~M Ljung.
\newblock {\em Time series analysis: forecasting and control}.
\newblock John Wiley \& Sons, 2015.

\bibitem{holt2004forecasting}
Charles~C Holt.
\newblock Forecasting seasonals and trends by exponentially weighted moving
  averages.
\newblock {\em International journal of forecasting}, 20(1):5--10, 2004.

\bibitem{su2022dlinear}
Chaoqun Su.
\newblock Dlinear makes efficient long-term predictions.
\newblock In {\em Proceedings of ACM Conference (Baidu KDD Cup)}, 2022.

\bibitem{achiam2023gpt}
Josh Achiam, Steven Adler, Sandhini Agarwal, Lama Ahmad, Ilge Akkaya,
  Florencia~Leoni Aleman, Diogo Almeida, Janko Altenschmidt, Sam Altman,
  Shyamal Anadkat, et~al.
\newblock Gpt-4 technical report.
\newblock {\em arXiv preprint arXiv:2303.08774}, 2023.

\bibitem{grattafiori2024llama}
Aaron Grattafiori, Abhimanyu Dubey, Abhinav Jauhri, Abhinav Pandey, Abhishek
  Kadian, Ahmad Al-Dahle, Aiesha Letman, Akhil Mathur, Alan Schelten, Alex
  Vaughan, et~al.
\newblock The llama 3 herd of models.
\newblock {\em arXiv preprint arXiv:2407.21783}, 2024.

\bibitem{radford2023robust}
Alec Radford, Jong~Wook Kim, Tao Xu, Greg Brockman, Christine Mcleavey, and
  Ilya Sutskever.
\newblock Robust speech recognition via large-scale weak supervision.
\newblock In Andreas Krause, Emma Brunskill, Kyunghyun Cho, Barbara Engelhardt,
  Sivan Sabato, and Jonathan Scarlett, editors, {\em Proceedings of the 40th
  International Conference on Machine Learning}, volume 202 of {\em Proceedings
  of Machine Learning Research}, pages 28492--28518. PMLR, 23--29 Jul 2023.

\bibitem{barrault2023seamlessm4t}
Lo{\"\i}c Barrault, Yu-An Chung, Mariano~Cora Meglioli, David Dale, Ning Dong,
  Paul-Ambroise Duquenne, Hady Elsahar, Hongyu Gong, Kevin Heffernan, John
  Hoffman, et~al.
\newblock Seamlessm4t: Massively multilingual \& multimodal machine
  translation.
\newblock {\em arXiv preprint arXiv:2308.11596}, 2023.

\bibitem{peebles2023scalable}
William Peebles and Saining Xie.
\newblock Scalable diffusion models with transformers.
\newblock In {\em Proceedings of the IEEE/CVF International Conference on
  Computer Vision (ICCV)}, pages 4195--4205, October 2023.

\bibitem{kirillov2023segment}
Alexander Kirillov, Eric Mintun, Nikhila Ravi, Hanzi Mao, Chloe Rolland, Laura
  Gustafson, Tete Xiao, Spencer Whitehead, Alexander~C. Berg, Wan-Yen Lo, Piotr
  Dollar, and Ross Girshick.
\newblock Segment anything.
\newblock In {\em Proceedings of the IEEE/CVF International Conference on
  Computer Vision (ICCV)}, pages 4015--4026, October 2023.

\bibitem{zhou2021informer}
Haoyi Zhou, Shanghang Zhang, Jieqi Peng, Shuai Zhang, Jianxin Li, Hui Xiong,
  and Wancai Zhang.
\newblock Informer: Beyond efficient transformer for long sequence time-series
  forecasting.
\newblock volume~35, pages 11106--11115, May 2021.

\bibitem{wu2021autoformer}
Haixu Wu, Jiehui Xu, Jianmin Wang, and Mingsheng Long.
\newblock Autoformer: Decomposition transformers with auto-correlation for
  long-term series forecasting.
\newblock In M.~Ranzato, A.~Beygelzimer, Y.~Dauphin, P.S. Liang, and J.~Wortman
  Vaughan, editors, {\em Advances in Neural Information Processing Systems},
  volume~34, pages 22419--22430. Curran Associates, Inc., 2021.

\bibitem{nie2022time}
Yuqi Nie, Nam~H Nguyen, Phanwadee Sinthong, and Jayant Kalagnanam.
\newblock A time series is worth 64 words: Long-term forecasting with
  transformers.
\newblock In {\em The Eleventh International Conference on Learning
  Representations}, 2023.

\bibitem{garza2023timegpt}
Azul Garza, Cristian Challu, and Max Mergenthaler-Canseco.
\newblock Timegpt-1.
\newblock {\em arXiv preprint arXiv:2310.03589}, 2023.

\bibitem{ansari2024chronos}
Abdul~Fatir Ansari, Lorenzo Stella, Ali~Caner Turkmen, Xiyuan Zhang, Pedro
  Mercado, Huibin Shen, Oleksandr Shchur, Syama~Sundar Rangapuram,
  Sebastian~Pineda Arango, Shubham Kapoor, Jasper Zschiegner, Danielle~C.
  Maddix, Hao Wang, Michael~W. Mahoney, Kari Torkkola, Andrew~Gordon Wilson,
  Michael Bohlke-Schneider, and Bernie Wang.
\newblock Chronos: Learning the language of time series.
\newblock {\em Transactions on Machine Learning Research}, 2024.
\newblock Expert Certification.

\bibitem{woo2024unified}
Gerald Woo, Chenghao Liu, Akshat Kumar, Caiming Xiong, Silvio Savarese, and
  Doyen Sahoo.
\newblock Unified training of universal time series forecasting transformers.
\newblock In {\em Forty-first International Conference on Machine Learning},
  2024.

\bibitem{liu2021time}
Minhao LIU, Ailing Zeng, Muxi Chen, Zhijian Xu, Qiuxia LAI, Lingna Ma, and
  Qiang Xu.
\newblock Scinet: Time series modeling and forecasting with sample convolution
  and interaction.
\newblock In S.~Koyejo, S.~Mohamed, A.~Agarwal, D.~Belgrave, K.~Cho, and A.~Oh,
  editors, {\em Advances in Neural Information Processing Systems}, volume~35,
  pages 5816--5828. Curran Associates, Inc., 2022.

\bibitem{wang2023micn}
Huiqiang Wang, Jian Peng, Feihu Huang, Jince Wang, Junhui Chen, and Yifei Xiao.
\newblock {MICN}: Multi-scale local and global context modeling for long-term
  series forecasting.
\newblock In {\em The Eleventh International Conference on Learning
  Representations}, 2023.

\bibitem{luo2024moderntcn}
Luo donghao and wang xue.
\newblock Modern{TCN}: A modern pure convolution structure for general time
  series analysis.
\newblock In {\em The Twelfth International Conference on Learning
  Representations}, 2024.

\bibitem{zeng2023transformers}
Ailing Zeng, Muxi Chen, Lei Zhang, and Qiang Xu.
\newblock Are transformers effective for time series forecasting?
\newblock volume~37, pages 11121--11128, Jun. 2023.

\bibitem{zhou2025transformers}
Yufa Zhou, Yixiao Wang, Surbhi Goel, and Anru~R Zhang.
\newblock Why do transformers fail to forecast time series in-context?
\newblock {\em arXiv preprint arXiv:2510.09776}, 2025.

\bibitem{wen2022transformers}
Qingsong Wen, Tian Zhou, Chaoli Zhang, Weiqi Chen, Ziqing Ma, Junchi Yan, and
  Liang Sun.
\newblock Transformers in time series: A survey.
\newblock {\em arXiv preprint arXiv:2202.07125}, 2022.

\bibitem{xiong2024efficient}
Yuwen Xiong, Zhiqi Li, Yuntao Chen, Feng Wang, Xizhou Zhu, Jiapeng Luo, Wenhai
  Wang, Tong Lu, Hongsheng Li, Yu~Qiao, Lewei Lu, Jie Zhou, and Jifeng Dai.
\newblock Efficient deformable convnets: Rethinking dynamic and sparse operator
  for vision applications.
\newblock {\em 2024 IEEE/CVF Conference on Computer Vision and Pattern
  Recognition (CVPR)}, pages 5652--5661, 2024.

\bibitem{wang2025uniconvnet}
Yuhao Wang and Wei Xi.
\newblock Uniconvnet: Expanding effective receptive field while maintaining
  asymptotically gaussian distribution for convnets of any scale.
\newblock {\em ArXiv}, abs/2508.09000, 2025.

\bibitem{makridakis2018m4}
Spyros Makridakis.
\newblock M4 dataset.
\newblock
  \url{https://github.com/M4Competition/M4-methods/tree/master/Dataset}, 2018.

\bibitem{liu2025semantic}
Hao Liu, Chun Yang, Xiaoxing Zhang, and Xiaobin Zhu.
\newblock Semantic-enhanced time-series forecasting via large language models.
\newblock {\em ArXiv}, abs/2508.07697, 2025.

\bibitem{wang2024timemixer++}
Shiyu Wang, Jiawei LI, Xiaoming Shi, Zhou Ye, Baichuan Mo, Wenze Lin,
  Ju~Shengtong, Zhixuan Chu, and Ming Jin.
\newblock Timemixer++: A general time series pattern machine for universal
  predictive analysis.
\newblock In {\em The Thirteenth International Conference on Learning
  Representations}, 2025.

\bibitem{hu2025context}
Yuxiao Hu, Qian Li, Dongxiao Zhang, Jinyue Yan, and Yuntian Chen.
\newblock Context-alignment: Activating and enhancing llm capabilities in time
  series.
\newblock {\em arXiv preprint arXiv:2501.03747}, 2025.

\bibitem{zhong2025time}
Siru Zhong, Weilin Ruan, Ming Jin, Huan Li, Qingsong Wen, and Yuxuan Liang.
\newblock Time-{VLM}: Exploring multimodal vision-language models for augmented
  time series forecasting.
\newblock In {\em Forty-second International Conference on Machine Learning},
  2025.

\bibitem{liu2024autotimes}
Yong Liu, Guo Qin, Xiangdong Huang, Jianmin Wang, and Mingsheng Long.
\newblock Autotimes: Autoregressive time series forecasters via large language
  models.
\newblock {\em Advances in Neural Information Processing Systems},
  37:122154--122184, 2024.

\bibitem{pan2024textbf}
Zijie Pan, Yushan Jiang, Sahil Garg, Anderson Schneider, Yuriy Nevmyvaka, and
  Dongjin Song.
\newblock S2ip-llm: Semantic space informed prompt learning with llm for time
  series forecasting.
\newblock In {\em International Conference on Machine Learning}, 2024.

\bibitem{jin2023time}
Ming Jin, Shiyu Wang, Lintao Ma, Zhixuan Chu, James~Y. Zhang, Xiaoming Shi,
  Pin-Yu Chen, Yuxuan Liang, Yuan-Fang Li, Shirui Pan, and Qingsong Wen.
\newblock Time-{LLM}: Time series forecasting by reprogramming large language
  models.
\newblock In {\em The Twelfth International Conference on Learning
  Representations}, 2024.

\bibitem{zhou2023one}
Tian Zhou, Peisong Niu, Xue Wang, Liang Sun, and Rong Jin.
\newblock One fits all: Power general time series analysis by pretrained lm.
\newblock {\em Advances in Neural Information Processing Systems 36}, 2023.

\bibitem{liu2023itransformer}
Yong Liu, Tengge Hu, Haoran Zhang, Haixu Wu, Shiyu Wang, Lintao Ma, and
  Mingsheng Long.
\newblock itransformer: Inverted transformers are effective for time series
  forecasting.
\newblock In {\em The Twelfth International Conference on Learning
  Representations}, 2024.

\bibitem{su2019robust}
Ya~Su, Youjian Zhao, Chenhao Niu, Rong Liu, Wei Sun, and Dan Pei.
\newblock Robust anomaly detection for multivariate time series through
  stochastic recurrent neural network.
\newblock In {\em Proceedings of the 25th ACM SIGKDD international conference
  on knowledge discovery \& data mining}, pages 2828--2837, 2019.

\bibitem{mathur2016swat}
Aditya~P Mathur and Nils~Ole Tippenhauer.
\newblock Swat: A water treatment testbed for research and training on ics
  security.
\newblock In {\em 2016 international workshop on cyber-physical systems for
  smart water networks (CySWater)}, pages 31--36. IEEE, 2016.

\bibitem{abdulaal2021practical}
Ahmed Abdulaal, Zhuanghua Liu, and Tomer Lancewicki.
\newblock Practical approach to asynchronous multivariate time series anomaly
  detection and localization.
\newblock In {\em Proceedings of the 27th ACM SIGKDD conference on knowledge
  discovery \& data mining}, pages 2485--2494, 2021.

\bibitem{xu2021anomaly}
Jiehui Xu, Haixu Wu, Jianmin Wang, and Mingsheng Long.
\newblock Anomaly transformer: Time series anomaly detection with association
  discrepancy.
\newblock In {\em International Conference on Learning Representations}, 2022.

\bibitem{zhou2024kan}
Quan Zhou, Changhua Pei, Fei Sun, HanJing, Zhengwei Gao, Haiming Zhang, Gaogang
  Xie, Dan Pei, and Jianhui li.
\newblock {KAN}-{AD}: Time series anomaly detection with
  kolmogorov{\textendash}arnold networks.
\newblock In {\em Forty-second International Conference on Machine Learning},
  2025.

\bibitem{liu2022non}
Yong Liu, Haixu Wu, Jianmin Wang, and Mingsheng Long.
\newblock Non-stationary transformers: Exploring the stationarity in time
  series forecasting.
\newblock {\em Advances in neural information processing systems},
  35:9881--9893, 2022.

\bibitem{ren2019time}
Hansheng Ren, Bixiong Xu, Yujing Wang, Chao Yi, Congrui Huang, Xiaoyu Kou, Tony
  Xing, Mao Yang, Jie Tong, and Qi~Zhang.
\newblock Time-series anomaly detection service at microsoft.
\newblock In {\em Proceedings of the 25th ACM SIGKDD international conference
  on knowledge discovery \& data mining}, pages 3009--3017, 2019.

\bibitem{boniol2021sand}
Paul Boniol, John Paparrizos, Themis Palpanas, and Michael~J Franklin.
\newblock Sand: Streaming subsequence anomaly detection.
\newblock {\em Proc. VLDB Endow.}, 14(10):1717--1729, 2021.

\bibitem{tuli2022tranad}
Shreshth Tuli, Giuliano Casale, and Nicholas~R. Jennings.
\newblock Tranad: Deep transformer networks for anomaly detection in
  multivariate time series data.
\newblock {\em Proc. VLDB Endow.}, 15:1201--1214, 2022.

\bibitem{breunig2000lof}
Markus~M Breunig, Hans-Peter Kriegel, Raymond~T Ng, and J{\"o}rg Sander.
\newblock Lof: identifying density-based local outliers.
\newblock In {\em Proceedings of the 2000 ACM SIGMOD international conference
  on Management of data}, pages 93--104, 2000.

\bibitem{xu2023fits}
Zhijian Xu, Ailing Zeng, and Qiang Xu.
\newblock {FITS}: Modeling time series with \$10k\$ parameters.
\newblock In {\em The Twelfth International Conference on Learning
  Representations}, 2024.

\bibitem{wang2024revisiting}
Zexin Wang, Changhua Pei, Minghua Ma, Xin Wang, Zhihan Li, Dan Pei, Saravan
  Rajmohan, Dongmei Zhang, Qingwei Lin, Haiming Zhang, et~al.
\newblock Revisiting vae for unsupervised time series anomaly detection: A
  frequency perspective.
\newblock In {\em Proceedings of the ACM web conference 2024}, pages
  3096--3105, 2024.

\bibitem{malhotra2015long}
Pankaj Malhotra, Lovekesh Vig, Gautam~M. Shroff, and Puneet Agarwal.
\newblock Long short term memory networks for anomaly detection in time series.
\newblock In {\em The European Symposium on Artificial Neural Networks}, 2015.

\bibitem{liu2024kan}
Ziming Liu, Yixuan Wang, Sachin Vaidya, Fabian Ruehle, James Halverson, Marin
  Soljacic, Thomas~Y. Hou, and Max Tegmark.
\newblock {KAN}: Kolmogorov{\textendash}arnold networks.
\newblock In {\em The Thirteenth International Conference on Learning
  Representations}, 2025.

\bibitem{wu2021current}
Renjie Wu and Eamonn~J Keogh.
\newblock Current time series anomaly detection benchmarks are flawed and are
  creating the illusion of progress.
\newblock {\em IEEE transactions on knowledge and data engineering},
  35(3):2421--2429, 2021.

\bibitem{bagnall2018uea}
Anthony Bagnall, Hoang~Anh Dau, Jason Lines, Michael Flynn, James Large, Aaron
  Bostrom, Paul Southam, and Eamonn Keogh.
\newblock The uea multivariate time series classification archive, 2018.
\newblock {\em arXiv preprint arXiv:1811.00075}, 2018.

\bibitem{zerveas2021transformer}
George Zerveas, Srideepika Jayaraman, Dhaval Patel, Anuradha Bhamidipaty, and
  Carsten Eickhoff.
\newblock A transformer-based framework for multivariate time series
  representation learning.
\newblock In {\em Proceedings of the 27th ACM SIGKDD conference on knowledge
  discovery \& data mining}, pages 2114--2124, 2021.

\bibitem{jungmambasl}
Yoo-Min Jung and Leekyung Kim.
\newblock Mambasl: Exploring single-layer mamba for time series classification.
\newblock In {\em The Fourteenth International Conference on Learning
  Representations}, 2026.

\bibitem{ahamed2025tscmamba}
Md~Atik Ahamed and Qiang Cheng.
\newblock Tscmamba: Mamba meets multi-view learning for time series
  classification.
\newblock {\em Information Fusion}, 120:103079, 2025.

\bibitem{wen2025shedding}
Yunshi Wen, Tengfei Ma, Ronny Luss, Debarun Bhattacharjya, Achille Fokoue, and
  Anak~Agung Julius.
\newblock Shedding light on time series classification using interpretability
  gated networks.
\newblock In {\em The Thirteenth International Conference on Learning
  Representations}, 2025.

\bibitem{eldele2024tslanet}
Emadeldeen Eldele, Mohamed Ragab, Zhenghua Chen, Min Wu, and Xiaoli Li.
\newblock Tslanet: Rethinking transformers for time series representation
  learning.
\newblock In {\em International Conference on Machine Learning}, 2024.

\bibitem{zhou2022fedformer}
Tian Zhou, Ziqing Ma, Qingsong Wen, Xue Wang, Liang Sun, and Rong Jin.
\newblock {FED}former: Frequency enhanced decomposed transformer for long-term
  series forecasting.
\newblock In {\em Proceedings of the 39th International Conference on Machine
  Learning (ICML)}, volume 162, pages 27268--27286. PMLR, 2022.

\end{thebibliography}
		
	\end{CJK*}
\end{document}